\documentclass[sn-mathphys,Numbered]{sn-jnl}

\usepackage{graphicx}%
\usepackage{multirow}%
\usepackage{amsmath,amssymb,amsfonts}%
\usepackage{amsthm}%
\usepackage{mathrsfs}%
\usepackage[title]{appendix}%
\usepackage{xcolor}%
\usepackage{textcomp}%
\usepackage{manyfoot}%
\usepackage{booktabs}%
\usepackage{algorithm}%
\usepackage{algorithmicx}%
\usepackage{algpseudocode}%
\usepackage{listings}%

\theoremstyle{thmstyleone}%
\theoremstyle{thmstyletwo}%

\theoremstyle{thmstylethree}%

\usepackage{tabularx}

\usepackage{lineno}
\begin{document}

\title[Article Title]{Blind source separation in 3$^{\textrm {rd}}$ generation gravitational-wave detectors}


\author*[1]{\fnm{Francesca} \sur{Badaracco}}\email{francesca.badaracco@ge.infn.it}

\author[2]{\fnm{Biswajit} \sur{Banerjee}}

\author[2]{\fnm{Marica} \sur{Branchesi}}

\author[1]{\fnm{Andrea} \sur{Chincarini}}

\affil*[1]{\orgname{INFN, Sez. di Genova}, \orgaddress{\street{via Dodecaneso}, \city{Genova}, \postcode{16146}, \country{Italy}}}

\affil[2]{\orgname{GSSI}, \orgaddress{\street{Viale F. Crispi}, \city{L'Aquila}, \postcode{67100}, \country{Italy}}}


\abstract{Third generation and future upgrades of current gravitational-wave detectors will present exquisite sensitivities which will allow to detect a plethora of gravitational wave signals. Hence, a new problem to be solved arises: the detection and parameter estimation of overlapped signals. The problem of separating and identifying two signals that overlap in time, space or frequency is something well known in other fields (e.g. medicine  and telecommunication). Blind source separation techniques are all those methods that aim at separating two or more unknown signals. This article provides a methodological review of the most common blind source separation techniques and it analyses whether they can be successfully applied to overlapped gravitational wave signals or not, while comparing the limits and advantages of each method.}




\maketitle

\section{Introduction}
Third-generation interferometric gravitational-wave (GW) detectors like the Einstein Telescope (ET) \cite{ET}
and Cosmic Explorer (CE) \cite{CE} will reach exquisite sensitivities, which will entail the detection of a plethora of signals \cite{Branchesi2023,CE2021}. In particular, the binary black hole (BBH) coalescence detection rate in ET is expected to be $10^5$–$10^6$ per year, whereas the binary neutron star (BNS) coalescence detection rate is expected to be 10$^5$ per year \cite{ET,Branchesi2023, Banerjee2022, Ronchini2022, Iacovelli2022, Satya2022}. The exceptionally high sensitivity permits accessing lower frequencies with respect to the current detectors, and therefore signals will be observed for a longer time. BNS may last from hours to days and NS-BH up to hours, depending on the lowest frequency in the ET sensitivity band \cite{Regimbau2009,Nitz2021,Banerjee2022}, whereas BBHs last for a shorter time (a median of approximately 45\,s with f$_\mathrm{low}$ = 5\,Hz \cite{Samajdar2021}). This implies that, especially at lower frequencies, there will be signals which will overlap one another (see Fig. \ref{fig:overlapped}). The overlap will not necessarily entail a bias in the estimated parameters with GW pipelines. Indeed, it has been argued that if two signals merge with a sufficient time difference ($\Delta T_{\mathrm{merging}} > 2$\,s), the parameters can be estimated with negligible bias. However, there are cases where parameter estimation biases can occur. This can happen if GW signals overlap significantly in time and/or frequency (i.e. similar chirp masses) or if they have close SNR values (usually, only the louder signal is recovered) \cite{Pizzati2022, Himemoto2021, Antonelli2021, Relton2021, Samajdar2021, Relton2022}. Finally, in case of highly overlapped signals (similar merging time, frequency and SNR) the final posterior distributions vary chaotically depending on the relative phase of two overlapped signals. In these cases the overlapped signals can be mistaken with highly precessing ones \cite{Relton2021}. The problem of overlapped signals influences parameter estimation and thus, for example, impacts general relativity tests and population modeling. Hence, an effective method to disentangle them is essential, both in view of the upgrade of present GW detectors \cite{Virgo_nEXT, LIGO_Voyager} and in view of third-generation GW detectors \cite{ET,CE}. In literature many methods and their variants are presented, however, selecting one or more techniques appropriate for separating overlapped GW signals should be done taking into consideration the time and frequency characteristics of the signals. 

Here, we conduct a systematic review of the literature to identify the most common blind source separation methods, regardless of the application. Because most methods differ only by small procedural differences, we summarized them into classes to capture their main features. Then, we review each methodological class to assess their shortcomings with respect to signal and noise characteristics. The present review is not intended to be an exhaustive discussion of all the blind source separation techniques and their nuances. Indeed, blind source separation is a general framework encompassing a very large number of techniques and derived extensions, depending on the problem at the hand, in that every technique can be implemented with various degree of computational complexity and properties, depending on its application. This manuscript will focus on the most common techniques, leaving at the interested Readers to delve deeper into the subject using the references provided here, particularly Ref. \cite{comon2010handbook,Pan2021,Oja2001,Nordhausen2018}. The present work instead, is meant to provide a general introduction to the readers who are unfamiliar with the subject and who may find blind source separation techniques useful in advancing their own field. 

\section{Literature review} \label{sec:review}
We followed the PRISMA (Preferred Reporting Items for Systematic Reviews and Meta-Analyses) method to select journal articles (not proceedings) from the available literature. The search was first conducted searching for the "blind source separation" keyword on ArXiv, PubMed, Ieeexplore and GoogleScholar. References were selected when signal separation was performed. Finally, Scopus was used to perform a wider search to understand the most commonly used methods. Therefore, the keywords used were "blind source separation" and keywords relative to the methods. The search was performed using the title, abstract, and keywords relative to the papers present in the Scopus database (e.g: TITLE-ABS-KEY ( "blind source separation" ) AND TITLE-ABS-KEY ( "non-negative matrix factorization") AND (LIMIT-TO ( DOCTYPE , "ar" )) ). Based on these references, we found that the following methods have been commonly used (and sometimes a combination of them): Independent Component Analysis (ICA), non-negative matrix factorization (NMF), Second Order Blind Identification (SOBI), Joint Approximation Diagonalization of Eigenmatrices (JADE), Fractional Fourier Transform (FrFT), time-frequency masking, an Algorithm for Multiple Unknown Signal Extraction (AMUSE), and a plethora of machine learning methods, in particular neural networks. Results of the search are presented in Tab. \ref{tab:methods}. 

\section{Main methods used in literature}\label{sec:methods}
 Depending on the Author and on the method, the problem of blind source separation can be stated as follows:
  \begin{equation}\label{eq:BSS}
     \mathbf{x}(t) = \mathbf{A}\mathbf{s}(t) + \mathbf{n}(t),
 \end{equation}
 where $\mathbf{x}(t)$ is the observed mixture, a vector containing the M observations (taken with M sensors), $\mathbf{s}(t)$ is a vector containing the M original signals, $\mathbf{n}(t)$ is the vector containing the noise affecting the M measurements and $\mathbf{A}$ is the mixing matrix, containing the constant coefficients that weight the true signals in the observed mixture. The goal of blind source separation is to find $\mathbf{A}$ and $\mathbf{s}(t)$ given $\mathbf{x}(t)$, without any prior knowledge of the sources and - in some frameworks - of the sensors array. Other authors and methods state the problem in a noiseless framework such as:
 \begin{equation}\label{eq:noise-free_BSS}
     \mathbf{x}(t) = \mathbf{A}\mathbf{s'}(t)
 \end{equation}

 Treating the blind source separation problem in a noiseless framework 
 works well in two cases: when the sensor's self-noise is negligible, and when the noise is a "source noise", i.e. when it  is added to each independent component and not to the observed mixtures: $\mathbf{x}(t) = \mathbf{A}(\mathbf{s}(t) + \mathbf{n}(t))$. In the second case, after solving the blind source separation problem the original independent components will have to be estimated from the noisy ones.  

\begin{table}[]
    \centering
    \begin{tabular}{|c|c|}
         \hline
         Method &  Publications\\
         \hline
         \hline
         Independent Component Analysis & 4024\\
         \hline
         Neural Network & 938\\
         \hline
         Non-negative matrix factorization & 745\\
         \hline
         SOBI  & 244\\
         \hline
         Empirical Mode Decomposition & 242\\
         \hline
         Time-frequency masking / Fractional Fourier Transform & 100\\            
         \hline
         AMUSE & 57\\
         \hline
         Fractional Fourier Transform & 22\\
         \hline
        \end{tabular}
    \caption{Each method was searched on Scopus among the title, the abstract, and the keywords of each paper using the string "blind source separation" (forcing each of these words to be followed by the next) the logical operator "AND" and the name of the method as written in the table (forcing again each word to be followed by the next).}
    \label{tab:methods}
\end{table}

\subsection{Independent component Analysis}
 As shown in Tab. \ref{tab:methods}, ICA (or any of its variations) is the most commonly used technique for blind source separation. It has been applied in many different fields, such as mechanical failure \cite{Yuan2021, Uddin2021}, electrical engineering \cite{HuaiweiLiao2003, wu2021peak}, array signal processing \cite{Kim2022}, seismology \cite{bharadwaj2017deblending, lubo2019independent}, biomedical signals \cite{james2004independent}, speech recognition \cite{hsieh2009independent}, wireless communication systems \cite{uddin2015applications}, and as a tool for noise reduction in GW physics \cite{derosa2012improvement, morisaki2016toward, akutsu2020application}. ICA is often applied to various types of time series (e.g., electroencephalograms \cite{Liebisch2021}), but also to images (e.g., neuroimaging \cite{Wu2021}). ICA is quite versatile: it is possible to find many versions of it in the literature, as well as cases where it is used in combination with other methods. Therefore, it can be easily adapted to various problems.\\
Assuming that there are N sources and M sensors, ICA can, in principle, separate signals when N=M. If N$>$M or N$<$M, the problem is said to be \textit{underdetermined} or \textit{overdetermined}, respectively. For the underdetermined case, other assumptions are required, such as the sparseness of data \cite{Kim2004underdetrmined}. In ICA, the observed values of the source signals and of the mixture ones are considered to be samples of random variables instead of samples of a time series. Based on this, ICA makes only two assumptions: 1) the N source signals are \textit{statistically independent} and 2) they must have \textit{non-Gaussian} distributions (at most one Gaussian component is allowed) \cite{stone2004, Oja2001}. It turns out \cite{hyvarinen2000independent} that the non-Gaussian requirement is equivalent to independence. Therefore, ICA retrieves the matrix $\mathbf{A}$ by maximizing the non-Gaussianity of the signals $\mathbf{s}$. ICA algorithms differentiate themselves based on the choice of the measure of non-Gaussianity. One choice is \textit{kurtosis}, which is computationally easy but not a robust measure of the non-Gaussianity being it very sensitive to outliers. Other options are based on the concept of entropy/information content (e.g. negentropy and mutual information). Finally, there is also the possibility of using the maximum likelihood, which is equivalent to the maximization of entropy \cite{cardoso1997infomax}. The most common algorithms used to perform ICA are FastICA \cite{Hyvarinen1999}, Infomax \cite{Bell1995} and JADE \cite{Cardoso1993}.
To conclude, ICA has two ambiguities: 1) it is not possible to estimate the variances (energies) of the independent components, (indeed they are usually assumed to all have variance one); and 2) it is not possible to determine the order of the independent components.\\  
However, ICA is usually treated in a noiseless framework (Eq. \ref{eq:noise-free_BSS}); this is because estimating the mixing matrix seems to be quite difficult when noise is present, therefore, for most applications reducing the noise before applying ICA might be the best approach \cite{Oja2001}. However, some models that tackle the noisy ICA problem do exist \cite{HYVARINEN199849,hyvarinen1999noisyGaussian,Cao_2003}. ICA is also considered an extension of the Principal Component Analysis (PCA), which instead minimizes the correlations between the signals \cite{tharwat2020ICA}. However,  ICA is considered more powerful than PCA because independence is a stronger property with respect to uncorrelatedness \cite{Oja2001}. Readers who wish to delve deeper into the subject or explore some applications of ICA are encouraged to consult Ref. \cite{hyvarinen2000independent} and Ref. \cite{comon2010handbook}. These references provide practical guidance on how to tackle a source separation problem. \\

\subsection{Second-Order Separation}
Second-order separation methods are yet another type of blind source separation technique. They rely on the assumption that the data are uncorrelated. However, this section will focus on a particular algorithm: Second-Order Blind Identification (SOBI), which  enjoys widespread use.\\
SOBI is another blind source separation technique, even though quite similar to ICA, it finds applications in many fields such as electroencephalogram artifact elimination \cite{ZangenehSoroush2022}, functional Magnetic Resonance Imaging \cite{HWu2021}, structures modal identification \cite{Lakshmi2021, SadeghiEshkevari2020}, power network quality monitoring \cite{deOliveira2021}. \\
SOBI was developed in 1997 by Belouchrani et al. \cite{Belouchrani1997} to separate temporally correlated sources, and it is a generalization of the AMUSE algorithm \cite{Tong, Pan2021}. It is based on second-order statistics of the mixtures and even allows the separation of Gaussian signals. It can be exploited for low-SNR signals or to separate sources with small spectral differences \cite{Belouchrani1997}. Both AMUSE and SOBI are widely used to separate uncorrelated and weakly stationary time series \cite{Pan2021}. SOBI  assumes that the sources are uncorrelated and affected by white noise (measurement noise), it identifies the mixing matrix $\mathbf{A}$ by finding a unitary matrix $\mathbf{U}$ that jointly diagonalizes (i.e. simultaneously diagonalize) the covariance matrices (second-order statistic) of the array output with different time delays, therefore the original signals can be retrieved as: 
\begin{equation}
    \mathbf{s}(t) = \mathbf{U}^{H}\mathbf{W}\mathbf{x}(t)
\end{equation}
Where $\mathbf{U}^{H}$ is the transpose conjugate unitary matrix and  $\mathbf{W}$ is the matrix used to whiten the data (the whitening is useful to search for a unitary matrix $\mathbf{U}$ instead of the mixture matrix $\mathbf{A}$). This demonstration can be found in Ref. \cite{Belouchrani1997}. 
Considering that SOBI implies diagonalizing matrices, it works better for moderate data set sizes. However, SOBI generalizations exist that try to improve the method's limitations \cite{Pan2021}.\\
Readers seeking to gain a deeper understanding of the method’s implementation and its applications are encouraged to consult Chapter 7 of Ref. \cite{comon2010handbook} and Ref. \cite{Pan2021}, which provides a comprehensive review of all second-order separation methods found in the literature. 

\subsection{Non-negative matrix factorization}
Non-negative matrix factorization (NMF) (or any of its variations) is the third most used method (see Tab \ref{tab:methods}) and, as ICA, it is used in many applications with different kind of data such as: heart-lung sound separation \cite{Wang2023}, spectral separation \cite{Leplat2022,Yakimov2021}, oil and gas exploration \cite{Gu2021}, computed tomography images \cite{Wei2021}, polarized signals \cite{Flamant2020}, brain images \cite{HernndezVillegas2019, Song2020}, cultural heritage \cite{Lyu2020}, fetal-mother electrocardiogram  separation \cite{Gurve2020} wireless sensor networks \cite{He2016} and of course speech recognition \cite{Li2012} to name a few. NMF is mostly used on images or even on 3D data \cite{Laudadio2016} as well as audio signals, which can be transformed into non-negative data using their spectrograms.

NMF is a method to describe data in its elementary parts. Indeed, it factorizes  the data (represented as matrices) into two matrices with the constraint that their values must be non-negative. This constraint comes from the fact that in many physical problems, this comes naturally and that it helps to give a physical interpretation to the separated sources. Unlike ICA and SOBI, NFM does not require as many sensors as the signals to separate. Instead, it interprets the data as a matrix ($\mathbf{V}$) and approximately factorizes it into two other matrices that can be interpreted as a dictionary ($\mathbf{W}$) and a matrix with the activation values of the dictionary elements ($\mathbf{H}$). A simple example applied to audio signals (similar to the GW case) can be found in Ref. \cite{gillis2020nonnegative}, Sec. 1.3.4.  
\begin{equation}
\mathbf{V} = \mathbf{W H}    
\end{equation}
Again, the problem can be defined as a minimization of a defined distance $\mathcal{D}$:
\begin{equation}
    \min_{\mathbf{W}, \mathbf{H}}\mathcal{D}(\mathbf{V}, \mathbf{WH})
\end{equation}
The choice of $\mathcal{D}$ depends on assumptions regarding the noise statistic (e.g. the Frobenius norm of the residual can be chosen when the noise is Gaussian, independent and identically distributed \cite{gillis2020nonnegative}). In NMF, data can be reconstructed only via summation (and no subtraction) of non negative components (or parts) of the whole; therefore, NMF is a parts-based representation of the data, i.e. the data can be decomposed into its fundamental parts, such as notes composing a musical piece. These parts (e.g. musical notes) are used to reconstruct the original signal ($\mathbf{V}$) by summing them together depending on their activation times stored in matrix $\mathbf{H}$. The non-negative constraint and the fact that $\mathbf{W}\mathbf{H}$ represents a summation of parts implies to have many vanishing elements, so $\mathbf{H}$ and $\mathbf{W}$ are sparse matrices. The only assumption here is that the parts are non-negative; however, no assumption on their statistical dependence is made \cite{lee1999NMF}. The assumption that negative numbers are physically meaningless does not prevent the application of NMF to data containing negative values (such as audio signals); they only need to be transformed in a non-negative form  beforehand (e.g., time-frequency spectra). In the case of audio signals, this might be a problem when reconstructing the signals from their spectrograms, because in this case, the phase information is missing and techniques for its reconstruction are required \cite{Phase_rec_2021}.  \\
Conventional NMF works well on signals with a constant spectrum but fails for signals showing a frequency sweep, such as chirps.  Convolutive NMF can instead cope well with time-evolving source signals spectra \cite{o2006convolutive}. It works by assigning to each source a sequence of successive spectra (instead of just one) and a corresponding activation pattern over time. Finally, NMF is based on the separability assumption of the data matrix \cite{gillis2020nonnegative} and the presence of noise spoils this assumption, making NFM extremely sensitive to noise. \\
Readers who are keen to gain a deeper understanding of NMF are directed to Nicolas Gillis’ book \cite{gillis2020nonnegative}. This book provides detailed descriptions of NMF, its associated algorithms, and various applications. 
\subsection{Time-frequency approaches}
One technique often used to separate signals that do not overlap significantly in the time-frequency domain (i.e., every time the frequency cell is dominated by a single source) is the time-frequency masking technique, which can be constructed from various clues based on the analysis of temporal, spectral, or spatial features of the sources. The time-frequency masks can be blindly determined when two anechoic mixtures are available \cite{Yilmaz2004} or by Neural Networks \cite{Yu2016}.  \\  
Another example of a time-frequency approach to separate signals is the Fractional Fourier Transform (FrFT) technique which is employed in many fields ranging from medical images denoising \cite{Mustafi2013}, marine biology \cite{Zhang2021}, linear chirp analysis \cite{Capus2003}, radio occultation \cite{Gorbunov2022}, speech emotion recognition \cite{Huang2022} to seismology \cite{Tian2021}. \\
The FrFT can also be employed to separate the signals. Indeed, when signals overlap in time, but not in frequency (e.g., two sine waves at different frequencies), separation is straightforward; however, if the signals overlap also in frequency, it is necessary to resort to other methods. Cowel et al. \cite{cowell2010separation} demonstrated that is possible to use the FrFT to separate linear frequency-modulated signals overlapping in time and frequency. The FrFT is a generalization of the Fourier Transform: the signals are transformed into a fractional domain, neither a time nor a frequency domain, but an intermediate between these two. Moreover, FrFT can be interpreted as a rotation of the Wigner-Ville distribution of a signal \cite{cowell2010separation}, after which a filter can be applied to separate the signals. FrFT is considered the best way to process linear chirp signals \cite{tao2009short} and its implementation cost is comparable to that of the Fourier Transform. However, FrFT is limited by the fact that it cannot locate the individual frequencies (FrFT-frequencies) with respect to time; thus, Tao et al. introduced the Short-Time FrFT \cite{tao2009short}, which is better at locating the frequency content of a chirp. However, the FrFT method is a good way to pre-process data which need then to be blindly separated either by using a Neural Network \cite{Zhang2021} or by finding other ways to blindly apply a filter \cite{zalevsky2007fractional}.\\
Here we mentioned only a few time-frequencies approaches related to blind source separation problems found in the literature. For a more comprehensive discussion, readers are referred to Ref. \cite{boashash2015time}, which serves as an exhaustive guide to time-frequency signal processing, including also blind source separation.

\subsection{Machine learning}
Another possibility for dealing with blind source separation is machine learning, particularly neural networks \cite{Agrawal2023}. \\
Neural networks have been used for signal separation in many fields such as seismology \cite{Novoselov2022}, speech separation \cite{Pfeifenberger2022}, energy disaggregation techniques \cite{VirtsionisGkalinikis2022} and image separation \cite{Sun2020}. Here we will not focus further on neural networks because this is a very heterogeneous topic, which would require a review by itself (see for instance Ref. \cite{Agrawal2023}) and hence it is outside of the scope of this paper. However, we acknowledge that ML techniques are rapidly gaining interest, therefore we only cite that the ML/DL approach is commonly applied in blind source separation problems and/or feature extraction jointly with classical techniques, such as ICA, NMF \cite{WANG2023104180, Karhunen_1997_NN_for_ICA} and time-frequencies approach, or as a single technique \cite{Bermant2021}. The Neural Networks are usually employed after pre-processing the data with the aforementioned techniques or directly to integrate them (e.g. Deep NMF \cite{chen2022survey}).
The most commonly used types of neural networks for blind source separation are convolutional neural networks, generative adversarial networks, and recurrent neural networks \cite{Agrawal2023}. Some of these techniques can directly deal with time domain signals (unlike NMF) and with a single-channel measurement \cite{Nakamura_2021, Bermant2021}, while others deal with images or time-frequency representations \cite{Dai2021}.
\newline
\begin{table}[]
    \centering
    \begin{tabular}{|m{0.7cm}|m{3.0cm}|m{3.0cm}|m{3.0cm}|m{3.0cm}|}
    
         \hline
         \footnotesize{} &  \footnotesize{Data assumptions} & \footnotesize{Limitations} & \footnotesize{Noise} & \footnotesize{Computational cost}\\
         \hline
         \hline
         \footnotesize{ICA} & \footnotesize{Independent, non-Gaussian} & \footnotesize{Not possible to estimate the amplitude of the signals; Not possible to determine to which source the two independent components belong} & \footnotesize{ICA is usually treated in a noiseless framework except some works on noisy models \cite{HYVARINEN199849}. }& \footnotesize{FastICA: O(dN); d = n° of sources and N = n° of samples}\\
         \hline
         \footnotesize{NMF} &  \footnotesize{The element of the matrix factors are non-negative} & \footnotesize{Conventional NMF fails with time varying source spectra. It can be solved by adding a sparseness constraint (Convolutional NMF). Moreover, the audio signal phases information is missing when the signal is reconstructed from the spectrogram of the sources} & \footnotesize{Extremely sensitive to the noise that makes the NFM problem near-separable instead of separable}  & \footnotesize{NP-hard problem}\\
         \hline
         \footnotesize{SOBI} & \footnotesize{Uncorrelated weakly stationary processes where most information is in the second moments. Constant mixing matrix} & \footnotesize{It may fail when signal higher moments contains crucial information} \cite{Pan2021} & \footnotesize{Suitable for low SNR data} & \footnotesize{O(d$^4$M); d = n° of sources and M = n° of covariance matrices}\\
         \hline
         \footnotesize{FrFT} & \footnotesize{Linear Frequency modulated signals} & \footnotesize{Probably less effective on non-linear chirps or even on chirps crossing each other in the time-frequency domain. Working on the spectrogram of the signal this method loses information on the phase } & \footnotesize{FrFT does not change performances with lower SNR, but the subsequent masking is certainly negatively affected by a lower SNR}&\footnotesize{O(N$^2$log$_2$N); N = n° of samples in the window}\\
         \hline
         \footnotesize{NN} & \footnotesize{None} & \footnotesize{It needs a large training dataset} &\footnotesize{NN robustness to noise depends on the training dataset (if it contains noise and how much)} & \footnotesize{
         It depends on the architecture}\\
         \hline
         \hline
        \end{tabular}
    \caption{}
    \label{tab:methods_char}
\end{table}

\section{Blind source separation for gravitational-wave data}\label{sec:GWsep}
\begin{figure}
    \centering
     \includegraphics[width=0.47\textwidth]{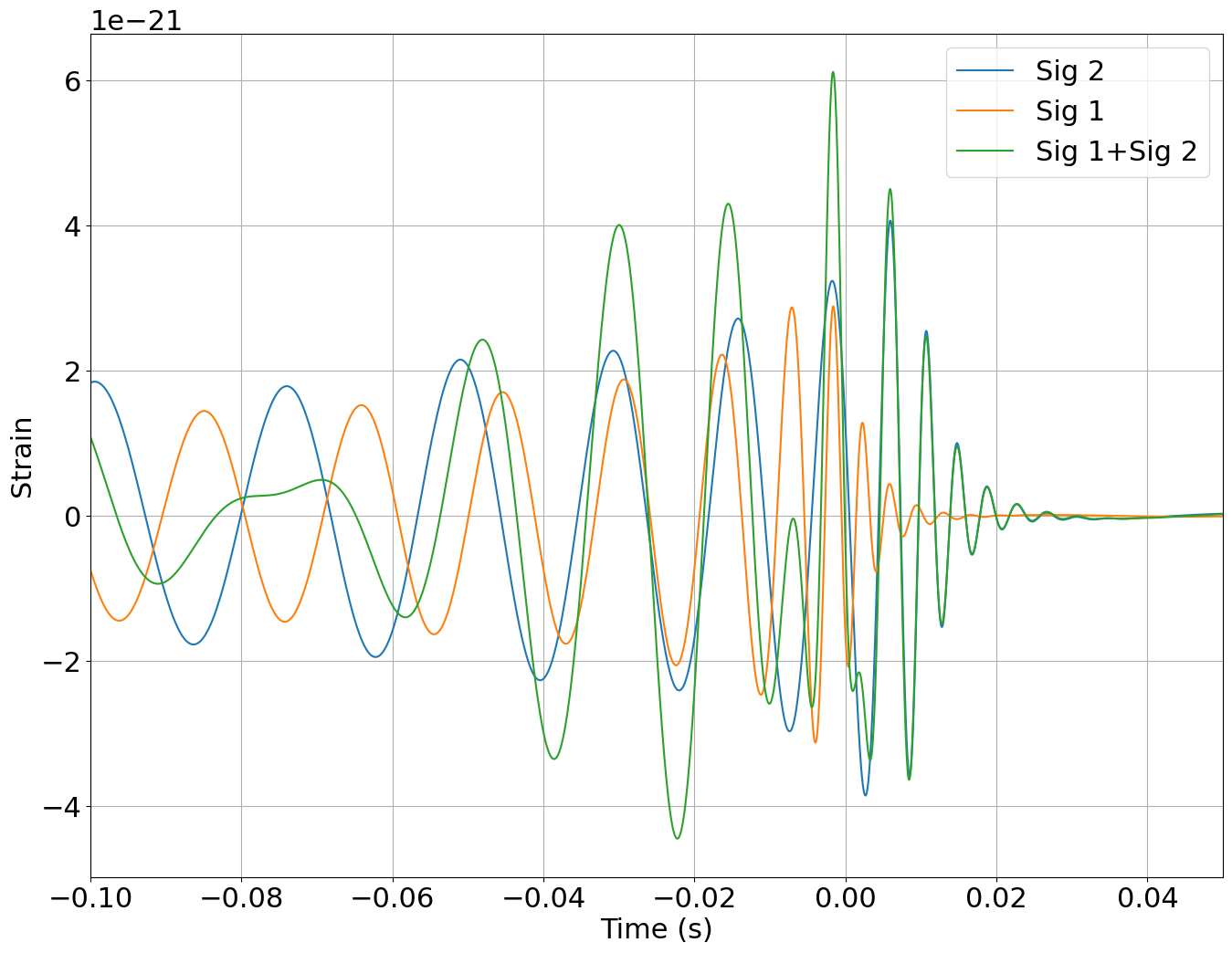}
     \includegraphics[width=0.47\textwidth]{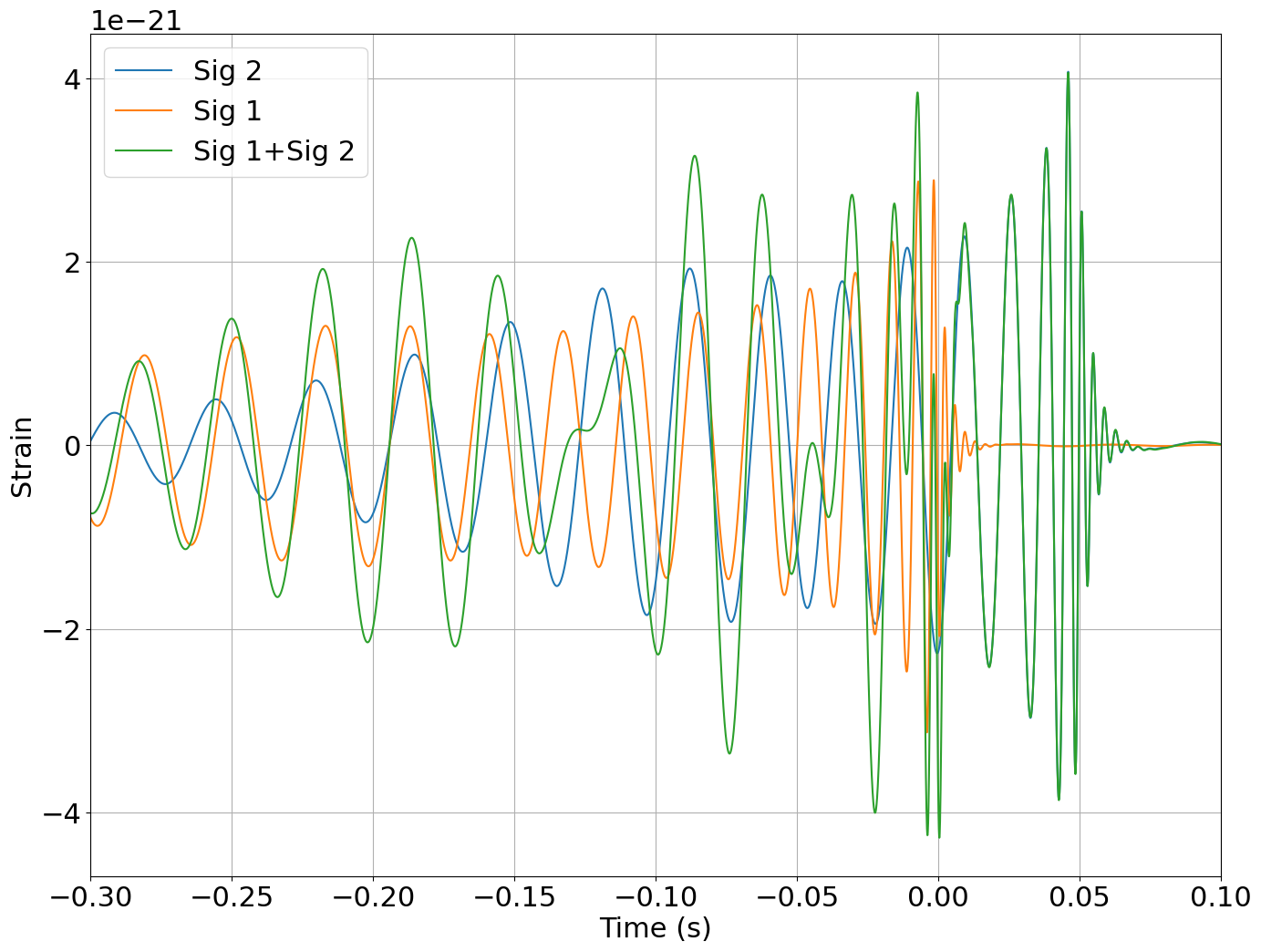}
     \includegraphics[width=0.47\textwidth]{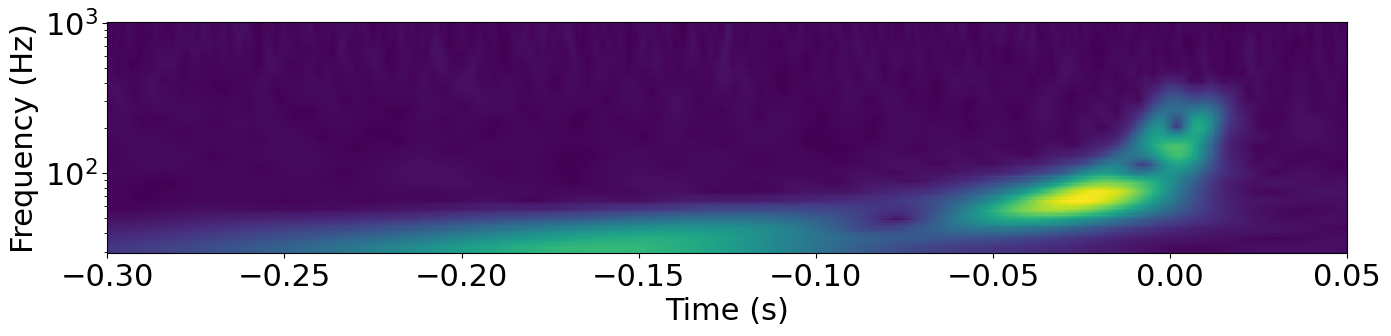}
     \includegraphics[width=0.47\textwidth]{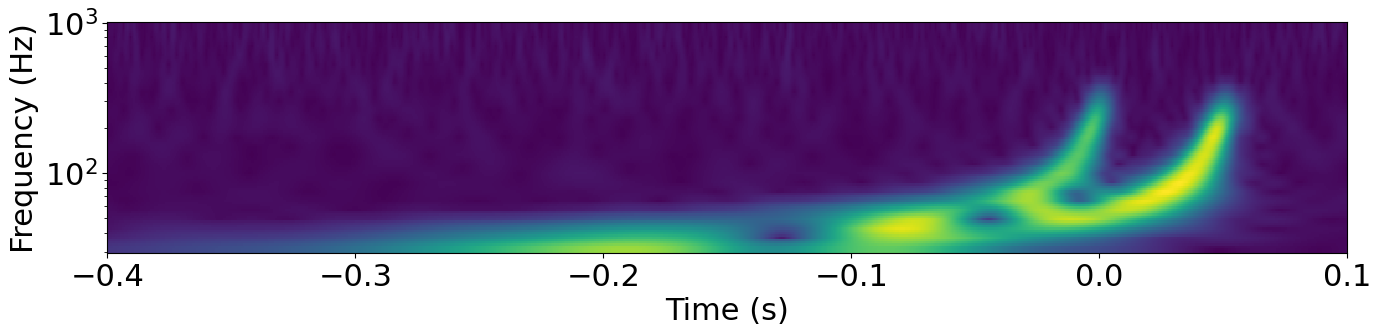}
     \centering
     \includegraphics[width=0.47\textwidth]{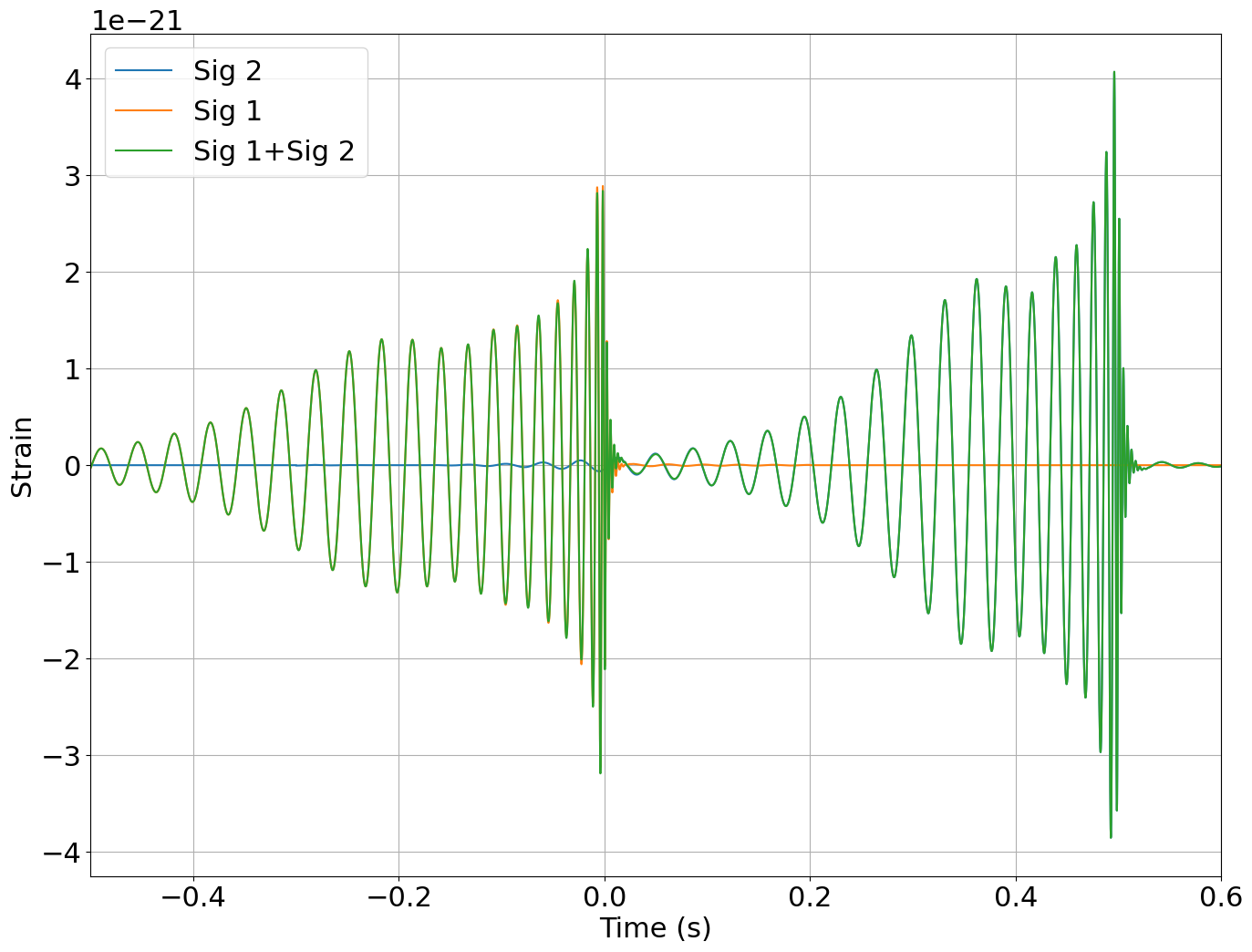}\\
     \includegraphics[width=0.47\textwidth]{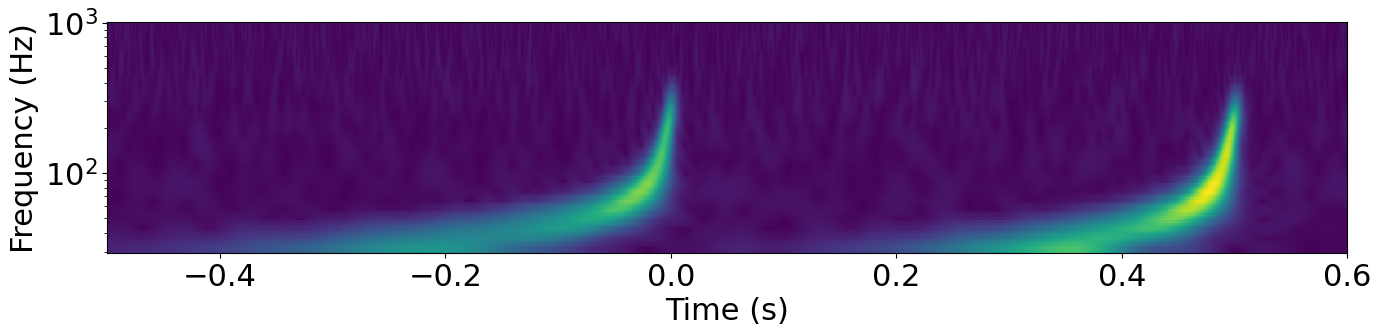}
     \caption{Overlapped GW signals example: time series and relative time domain representation (a Q-transform \cite{Q_transform} was applied). Plots made with pyCBC \cite{pyCBC}.\textit{Top left}: the time delay between the mergers is 0.01\,s, \textit{Top right}: the time delay between the mergers is 0.05\,s, \textit{Bottom}: the time delay between the mergers is 0.5\,s.}
    \label{fig:overlapped}
\end{figure}

Gravitational wave signals can be of different types: transient signals from compact binary coalescence (CBC), short transient unmodeled signals (burst) from sources such as core-collapse of massive stars and continuous signals from isolated neutron stars. CBC will be the most numerous signal, and they will overlap, particularly in the low-frequency band. Therefore, studies of overlapping signals mostly focuses on compact binaries  \cite{Relton2022,Pizzati2022,Himemoto2021,Antonelli2021,Relton2021,Samajdar2021}. However, it is possible that glitches (i.e. non-Gaussian transients) or some type of bursts might overlap to CBC signals. Hence, it is important to have a method to separate overlapped signals, possibly independently from the signal family (CBC or other). \\
The blind source separation of GW signals has some differences with respect to a classical cocktail party problem, such as that of separating speakers in a noisy room using microphones.
Firstly, the number of available sensors is limited and not always the same. ET is designed to be constituted by three detectors, but, due to their limited duty cycle, there will likely be periods in which only two, or even one, will be in observing mode. This greatly reduces the possibility of effectively disentangling signals (many techniques require the number of sensors equal to the number of sources). Moreover, while the SNR of two speakers talking in a noisy room can be assumed to be similar, the SNR of two GW overlapped signals might be quite different, and given that GW signals are very feeble, we can expect many low-SNR signals as well. Some techniques are usually treated in a noiseless framework, which in the GW case is not true: the noises in GW detectors are not negligible and do not follow a Gaussian statistic. However, some of the noise sources in the triangular ET detector will be common to all three instruments, therefore their noises will not be independent. 
Another issue is that the spectral content varies in time, and this complicates further the problem (e.g. the classic NMF does not cope well with time-varying sources).   
Finally, the GW case has one noteworthy advantage compared to many other cocktail party problems: the detector noise and a large portion of the source signals can be accurately modelled (see Figures \ref{fig:overlapped} and \ref{fig:CBC}), thus helping in the design of the blind source separation problem or even allowing to train a Neural Network on a large train data set.

\subsection{GW data characterization in ET}\label{sec:GWsep}
GW detector output time series, $h(t)$ is assumed to have the form:
\begin{equation} \label{eq:GW_strain}
    h(t) = s(t) + n(t)
\end{equation}
where $s(t)$ and $n(t)$ represent the astrophysical signal and the detector noise, respectively. In current GW detectors, $s(t)$ has always been considered as a single event, but in future GW detectors it will represent a sum of the detected signals.\\
Furthermore, each signal is projected differently onto each detector, therefore Eq. \ref{eq:GW_strain} should be represented as Eq. \ref{eq:BSS}, where $\mathbf{A}$ contains information about the different SNR of the signals, the source orientation with respect to the detectors and their antenna pattern. The vector $\mathbf{n(t)}$ would represent the noise of each detector. 
ET will detect a plethora of CBC signals, as well as bursts. The CBC signals are power-law chirps \cite{CHASSANDEMOTTIN1999252}. However, depending on the merger phase, their chirp behaviour can be different: during the early inspiralling phase they are almost monochromatic but closer to the merger the chirp is more evident. ET will be sensitive in a frequency band ranging from a few Hz to 10\,kHz. GW signals (mainly CBCs) in ET will overlap between themselves, but also with glitches: nonstationary, narrow-band and loud noise. 
\begin{figure}
    \centering
     \includegraphics[width=0.8\textwidth]{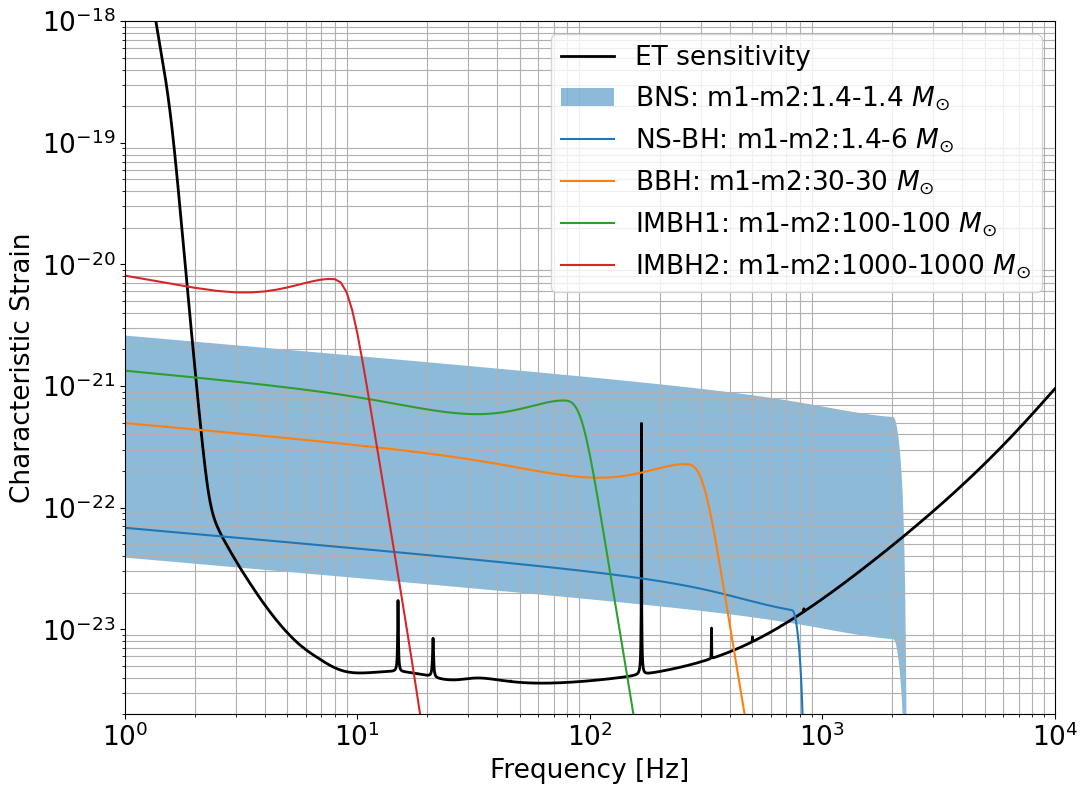}
     \caption{The black curve represents the estimated sensitivity curve of ET for a 10\,km xylophone configuration \cite{Branchesi2023}. For a comprehensive description of how this noise curve is recovered see Ref. \cite{ET}. The colored band represent the binary coalescence of binary neutron stars (BNS) ranging from distances of 100 Mpc to 6700\,Mpc, while solid lines represent neutron star-black hole (NS-BH), binary black holes (BBH) and intermediate mass black holes (IMBH) at 6700\,Mpc. The area between ET sensitivity and the signals is proportional to the relative SNR \cite{Moore2014}. The signal curves were plotted using the pyCBC software \cite{pyCBC} and the models IMRPhenomPv2\_NRTidal (NS-BH and BNS) and IMRPhenomD (BBH and IMBH) were used.}
    \label{fig:CBC}
\end{figure}
There will be pipelines dedicated to the removal of such glitches, as in Virgo-LIGO \cite{Davis2022}; however, a potential blind signal separation pipeline could help in separating a glitch from a real GW signal. Signals in ET will be seen with SNR varying from 10 to 10$^{\mathrm{3}}$, depending on the distances and masses involved and on the signal permanence in the detector band \cite{Maggiore2020,Branchesi2023}. Therefore, a blind source separation algorithm should be able to work with signals with different SNR. For example, Fig. \ref{fig:CBC} illustrates a comparison between various CBC signals and the sensitivity curve of ET. The loudness of the signal varies based on the distance, while the frequency shifts higher or lower depending on the masses involved.

\subsection{Overlapped signals in LISA}\label{sec:LISA}
The Laser Interferometric Space Antenna (LISA) will be a space-borne GW detector sensitive in the mHz band where it will detect thousands of overlapping sources that will form the so-called "source confusion" \cite{Crowder2004}. It is therefore natural to think that the techniques developed by the LISA community to address source confusion might work for ET as well. However, we must consider that there are a few fundamental differences between the signals detected by LISA and ET. \\
The LISA will detect galactic ultracompact binaries (mostly white dwarfs). In the LISA sensitivity band, these sources will appear as continuous and nearly monochromatic; therefore, the signal will not significantly evolve in frequency (even though it will have a frequency modulation due to the LISA revolution ) \cite{Littenberg2020}. The LISA community solution to this problem is to employ a global fit that seeks a global solution that simultaneously fits all the  signals in the band.\\
In ET, mergers will occur in the sensitivity band, which implies that these sources will be present in the band for much shorter times than in LISA. Moreover, the pipelines for detection will need to be very fast to find these signals to ensure multi-messenger astronomy.
\subsection{Current GW search pipelines} \label{sec:pipline}
Pipelines dedicated to detect GW signals can be divided in two groups: modelled and unmodelled searches. Modelled pipelines search for known signals from Binary Neutron Star (BNS), Binary Black Holes (BBH) and Neutron Star-Black Hole mergers (NS-BH). Unmodelled pipelines are more versatile in that they can spot signal of any type (e.g. core-collapse supernovae or unknown and exotic signals).\\
Modelled pipelines are: GstLAL \cite{GstLAL}, MBTA \cite{MBTA}, PyCBC Live \cite{PyCBC_LIVE} and SPIIR \cite{SPIIR}. They are matched-filter based, this means that they rely on discrete bank of templates containing signals covering a wide parameter space. Unmodelled searches are: cWB \cite{cWB} and oLIB \cite{oLIB} and BayesWave \cite{BayesWave} which is a follow-up of cWB triggers. They search for coincident excesses of the SNR in the GW detectors. Current pipelines assume that only a signal at a time is present. This assumption will break down in upgraded GW interferometers as well as in third generation GW detectors. Relton et al. \cite{Relton2022} studied how well modelled (pyCBC) and unmodelled (cWB) pipelines can perform in detecting these signals. The Authors conclude that the two pipelines might be modified and used in combination to address overlapping signals, at least in a majority of overlapping configurations. Surely, unmodelled pipeline searches that provide a signal reconstruction, like cWB, can be considered as a kind of blind source separation technique.
\section{Expectations of blind source separation techniques on GW data}\label{sec:discussion}
In ET, the so-called confusion noise will mostly be produced by CBC signals, therefore the methods selected to perform blind source separation will have to be effective, especially for these types of signals. In the following, the advantages and disadvantages of the methods described in Sec. \ref{sec:methods} will be discussed concerning the specific issue of the overlapped sources in ET. Moreover, Tab. \ref{tab:methods_char} shows a summary of the main advantages and disadvantages of the various methods. 
\newline
\newline
\textit{Independent Component Analysis and SOBI}\\
Conventional ICA and SOBI can separate N sources when an equal number of sensors is available. ET will comprise three nested interferometers arranged in a triangular shape \cite{ET}; therefore, it would be possible to disentangle from two to three overlapped signals, depending on the number of operating interferometers at the time of the signal arrival (n.b. one more sensor can be added when considering CE). However, it will not be possible to know the exact number of signals present in the data and, most of the time, only one or two ET interferometers will be in the observing mode. This implies that the unknown presence of a larger number of signals in the data stream with respect to the number of observing interferometers might lead to poor signal separation (e.g., the separated signals might still be a mixture of the additional signals present in the data).  \\
Another minor issue related to both ICA and SOBI is that they would fail to disentangle lensed GW signals because they would simply be the same signal arriving at different times (i.e., they would be correlated).\\
Both ICA and SOBI do not make particularly restrictive assumptions about the signals; however, SOBI assumes the signals to be uncorrelated and weakly stationary (see Tab. \ref{tab:methods_char}) and common GW signals cannot be considered stationary at any degree (except for the various kinds of stochastic GW backgrounds). Thus, ICA would be more appropriate than SOBI to disentangle CBCs and other types of signals beside them.
Finally, another important difference between ICA and SOBI  is that ICA is usually considered in a noiseless framework, whereas SOBI is indicated for signals with a low SNR, which for some CBC signals will be a concrete
possibility in ET. Therefore, it would be appropriate to better inspect noisy ICA, although classical ICA can be considered as a denoising tool (see Ref. \cite{derosa2012improvement, Forte2011WignerVille} ).
\newline
\newline
\textit{NMF and Time-frequency approaches}\\
NMF and time-frequency approaches are intimately connected when concerning the analysis of GW data; indeed, NMF assumes non-negative data and this implies that GW data must be analyzed in a time-frequency representation. This also means that the computational complexity of the NMF algorithm should allow to take into account the time-frequency transformation. Another aspect to bear in mind about NMF application to GW data is that it fails to deal with time-varying source spectra, such as GW chirp signals. Nevertheless, this problem can be bypassed by adding a sparseness constraint (e.g., convolutional NMF) \cite{o2006convolutive}. It is therefore possible that a convolutional NMF plus some other tricks might be successful in separating GW chirp signals in ET. Moreover, being that CBC are power-law chirps \cite{CHASSANDEMOTTIN1999252}, methods that work well on linear chirps, such as FrFT, are  in principle less effective on CBC signals. Additionally, using a time-frequency representation of the data leads to the loss of the phase information, which still is not a limiting factor since the phase information can be stored beforehand and later used to reconstruct the time-domain signals. \\
The possible uses of the Wigner-Ville or Hilbert-Huan transforms as a data pre-processing tool is also worth mentioning. For example, the Hilbert-Huan transform \cite{Huang2005} can be used to obtain a time-frequency representation of GW data to better highlight the instantaneous frequency behavior of signals \cite{Hu2022} and it could be used for NMF. The Wigner-Ville transform \cite{Debnath2002} can provide a high resolution both in time and frequency, and it is optimal for detecting linear chirps (hence, it is probably less effective on power-law chirps such as CBCs) \cite{CHASSANDEMOTTIN1999252}. As the Hilbert-Huang transform, it can be used as a preprocessing tool for NMF. Time-frequency approaches can be implemented as pre-processing for NMF, but they can also be used in combination with a neural network to blindly construct the opportune time-frequency mask that can be used to separate the signals.
\newline
\newline
\textit{Neural Networks}\\
GW signals have the great advantage that extremely detailed models of both the detector noise and a large portion of the signals are available. This makes of them natural candidates for neural network applications, which might become another possible approach to the blind source separation problem in the GW field. Indeed, it would be  possible to train neural networks on a large synthetic data set with artificially injected overlapped GW signals and then test it on another synthetic GW data set.    However, this method has the limitation that while it might be extremely effective in disentangling BBHs, BNSs, and NS-BH signals, for which detailed models are available,  it would certainly fail for unmodelled sources that cannot be included in the training data set. Of course, it depends on the final goal: it is already known that current GW data analysis pipelines work quite well in 
the presence of overlapped signals; therefore, it might be possible to simply employ a neural network as an aid to current pipelines (e.g., when signals overlap with a very small difference in merger time). 
\newline
\newline  
Once the optimal approach to the blind source separation problem in ET will be identified, it will important to understand also the interactions between the pipelines searching for GW signals and those disentangling overlapped signals. Given the effectiveness of search pipelines, one way to deal with overlapped signals might be to take the signal found by a search pipeline and check if it can be represented as a mixture of signals or if it is most likely a single signal. In this manner, the blind source separation algorithm would not constitute a bottleneck for the search pipelines, but would instead work as an additional aid. Indeed, it could prompt a second parameter estimation when the signal is suspected to consist of two close-merging signals. However, if the blind source separation pipeline will be sufficiently fast to not constitute a bottleneck for search pipelines, it could be used before them. Indeed, the computational effort required by the blind source separation technique is very important: the faster the speed, the better, because a correct parameter estimation allows a better definition of the sky localization of the source and allows a fast electromagnetic follow-up.
\section{Conclusion} \label{sec:conclusion}
ET and CE will improve sensitivities with respect to current detectors in two ways: by pushing it down to 2-3\,Hz and by enhancing it by an order of magnitude, or more, over the entire bandwidth. This entails the detection of a huge number of signals, many of which will be overlapped. GW search pipelines seem to demonstrate of being capable of retrieving signals with negligible bias most of the time, but on the precision of the parameter estimation the debate is still open. However, there are cases in which the overlap is extremely important and the pipelines fail. For these cases, it is necessary to develop an aid to the pipelines in order to improve the parameter estimation of the overlapped signals. \\
In this paper, we reviewed the most common techniques to blindly separate overlapped signals. The intent was to understand whether some of these techniques (or a combination of them) could cope better with the GW overlapping problem. However, all these techniques appear to have some kind of constraints regarding the type of signal they can separate, and, in general, these constraints do not fit well with the GW signals characteristics (see Tab. \ref{tab:methods_char}). Therefore, it is likely that none of these techniques -- at least not in the vanilla implementation -- will perform better or faster than current GW search pipelines, which seems already able to tackle most overlapping configurations. However, we do not rule out the potential use of these techniques following further research and development to meet the specific requirements of GW data.
Moreover, further research is needed to adapt blind source separation techniques to GW pipelines in order to get optimal results, although it is not clear yet whether these are to be prepended, or embedded in the GW analysis code. Once validated, separation techniques could used post-analysis to check whether a seemingly single signal consists of more components. There might also be room for optimization by understanding which kind of signals are mostly misinterpreted by the pipelines.\\           
Such blind source separation techniques could also be exploited for an ET denoising by separating GW signals overlapped to glitches, as in the case of GW170817 \cite{Abbott2017}.   \\
To conclude, we would like to point out that these blind source separation techniques might also be exploited in other GW experiments, such as the Lunar Gravitational-Wave Antenna (LGWA \cite{Harms2021}), where the seismic lunar signal will need to be separated from the actual GW signal.

\bmhead{Acknowledgments}
F.B. was financially supported by INFN with the grant no. 24526 funded through the project
PRIN\_2020KB33TP.\\
This article is based upon work from COST Action CA17137, supported by COST (European Cooperation in Science and Technology).\\
F.B. thanks Marco Drago for the useful comments provided in section \ref{sec:pipline}.

\bibliography{Manuscript}


\begin{thebibliography}{116}
\ifx \bisbn   \undefined \def \bisbn  #1{ISBN #1}\fi
\ifx \binits  \undefined \def \binits#1{#1}\fi
\ifx \bauthor  \undefined \def \bauthor#1{#1}\fi
\ifx \batitle  \undefined \def \batitle#1{#1}\fi
\ifx \bjtitle  \undefined \def \bjtitle#1{#1}\fi
\ifx \bvolume  \undefined \def \bvolume#1{\textbf{#1}}\fi
\ifx \byear  \undefined \def \byear#1{#1}\fi
\ifx \bissue  \undefined \def \bissue#1{#1}\fi
\ifx \bfpage  \undefined \def \bfpage#1{#1}\fi
\ifx \blpage  \undefined \def \blpage #1{#1}\fi
\ifx \burl  \undefined \def \burl#1{\textsf{#1}}\fi
\ifx \doiurl  \undefined \def \doiurl#1{\url{https://doi.org/#1}}\fi
\ifx \betal  \undefined \def \betal{\textit{et al.}}\fi
\ifx \binstitute  \undefined \def \binstitute#1{#1}\fi
\ifx \binstitutionaled  \undefined \def \binstitutionaled#1{#1}\fi
\ifx \bctitle  \undefined \def \bctitle#1{#1}\fi
\ifx \beditor  \undefined \def \beditor#1{#1}\fi
\ifx \bpublisher  \undefined \def \bpublisher#1{#1}\fi
\ifx \bbtitle  \undefined \def \bbtitle#1{#1}\fi
\ifx \bedition  \undefined \def \bedition#1{#1}\fi
\ifx \bseriesno  \undefined \def \bseriesno#1{#1}\fi
\ifx \blocation  \undefined \def \blocation#1{#1}\fi
\ifx \bsertitle  \undefined \def \bsertitle#1{#1}\fi
\ifx \bsnm \undefined \def \bsnm#1{#1}\fi
\ifx \bsuffix \undefined \def \bsuffix#1{#1}\fi
\ifx \bparticle \undefined \def \bparticle#1{#1}\fi
\ifx \barticle \undefined \def \barticle#1{#1}\fi
\bibcommenthead
\ifx \bconfdate \undefined \def \bconfdate #1{#1}\fi
\ifx \botherref \undefined \def \botherref #1{#1}\fi
\ifx \url \undefined \def \url#1{\textsf{#1}}\fi
\ifx \bchapter \undefined \def \bchapter#1{#1}\fi
\ifx \bbook \undefined \def \bbook#1{#1}\fi
\ifx \bcomment \undefined \def \bcomment#1{#1}\fi
\ifx \oauthor \undefined \def \oauthor#1{#1}\fi
\ifx \citeauthoryear \undefined \def \citeauthoryear#1{#1}\fi
\ifx \endbibitem  \undefined \def \endbibitem {}\fi
\ifx \bconflocation  \undefined \def \bconflocation#1{#1}\fi
\ifx \arxivurl  \undefined \def \arxivurl#1{\textsf{#1}}\fi
\csname PreBibitemsHook\endcsname

\bibitem[\protect\citeauthoryear{Team}{2020}]{ET}
\begin{botherref}
\oauthor{\bsnm{Team}, \binits{E.S.C.E.}}:
Design report update 2020 for the einstein telescope
(2020)
\end{botherref}
\endbibitem

\bibitem[\protect\citeauthoryear{Reitze et~al.}{2019}]{CE}
\begin{botherref}
\oauthor{\bsnm{Reitze}, \binits{D.}},
\oauthor{\bsnm{Adhikari}, \binits{R.X.}},
\oauthor{\bsnm{Ballmer}, \binits{S.}},
\oauthor{\bsnm{Barish}, \binits{B.}},
\oauthor{\bsnm{Barsotti}, \binits{L.}},
\oauthor{\bsnm{Billingsley}, \binits{G.}},
\oauthor{\bsnm{Brown}, \binits{D.A.}},
\oauthor{\bsnm{Chen}, \binits{Y.}},
\oauthor{\bsnm{Coyne}, \binits{D.}},
\oauthor{\bsnm{Eisenstein}, \binits{R.}}, et al.:
Cosmic explorer: the us contribution to gravitational-wave astronomy beyond
  ligo.
arXiv preprint arXiv:1907.04833
(2019)
\end{botherref}
\endbibitem

\bibitem[\protect\citeauthoryear{Branchesi et~al.}{2023}]{Branchesi2023}
\begin{barticle}
\bauthor{\bsnm{Branchesi}, \binits{M.}}, \betal:
\batitle{Science with the einstein telescope: a comparison of different
  designs}.
\bjtitle{Journal of Cosmology and Astroparticle Physics}
\bvolume{2023}(\bissue{07}),
\bfpage{068}
(\byear{2023})
\doiurl{10.1088/1475-7516/2023/07/068}
\end{barticle}
\endbibitem

\bibitem[\protect\citeauthoryear{{Evans} et~al.}{2021}]{CE2021}
\begin{botherref}
\oauthor{\bsnm{{Evans}}, \binits{M.}},
\oauthor{\bsnm{{Adhikari}}, \binits{R.X.}},
\oauthor{\bsnm{{Afle}}, \binits{C.}},
\oauthor{\bsnm{{Ballmer}}, \binits{S.W.}},
\oauthor{\bsnm{{Biscoveanu}}, \binits{S.}},
\oauthor{\bsnm{{Borhanian}}, \binits{S.}},
\oauthor{\bsnm{{Brown}}, \binits{D.A.}},
\oauthor{\bsnm{{Chen}}, \binits{Y.}},
\oauthor{\bsnm{{Eisenstein}}, \binits{R.}},
\oauthor{\bsnm{{Gruson}}, \binits{A.}},
\oauthor{\bsnm{{Gupta}}, \binits{A.}},
\oauthor{\bsnm{{Hall}}, \binits{E.D.}},
\oauthor{\bsnm{{Huxford}}, \binits{R.}},
\oauthor{\bsnm{{Kamai}}, \binits{B.}},
\oauthor{\bsnm{{Kashyap}}, \binits{R.}},
\oauthor{\bsnm{{Kissel}}, \binits{J.S.}},
\oauthor{\bsnm{{Kuns}}, \binits{K.}},
\oauthor{\bsnm{{Landry}}, \binits{P.}},
\oauthor{\bsnm{{Lenon}}, \binits{A.}},
\oauthor{\bsnm{{Lovelace}}, \binits{G.}},
\oauthor{\bsnm{{McCuller}}, \binits{L.}},
\oauthor{\bsnm{{Ng}}, \binits{K.K.Y.}},
\oauthor{\bsnm{{Nitz}}, \binits{A.H.}},
\oauthor{\bsnm{{Read}}, \binits{J.}},
\oauthor{\bsnm{{Sathyaprakash}}, \binits{B.S.}},
\oauthor{\bsnm{{Shoemaker}}, \binits{D.H.}},
\oauthor{\bsnm{{Slagmolen}}, \binits{B.J.J.}},
\oauthor{\bsnm{{Smith}}, \binits{J.R.}},
\oauthor{\bsnm{{Srivastava}}, \binits{V.}},
\oauthor{\bsnm{{Sun}}, \binits{L.}},
\oauthor{\bsnm{{Vitale}}, \binits{S.}},
\oauthor{\bsnm{{Weiss}}, \binits{R.}}:
{A Horizon Study for Cosmic Explorer: Science, Observatories, and Community}.
arXiv e-prints,
2109--09882
(2021)
\doiurl{10.48550/arXiv.2109.09882}
{\href{https://arxiv.org/abs/2109.09882}{{arXiv:2109.09882}}}
{[astro-ph.IM]}
\end{botherref}
\endbibitem

\bibitem[\protect\citeauthoryear{{Banerjee} et~al.}{2022}]{Banerjee2022}
\begin{botherref}
\oauthor{\bsnm{{Banerjee}}, \binits{B.}},
\oauthor{\bsnm{{Oganesyan}}, \binits{G.}},
\oauthor{\bsnm{{Branchesi}}, \binits{M.}},
\oauthor{\bsnm{{Dupletsa}}, \binits{U.}},
\oauthor{\bsnm{{Aharonian}}, \binits{F.}},
\oauthor{\bsnm{{Brighenti}}, \binits{F.}},
\oauthor{\bsnm{{Goncharov}}, \binits{B.}},
\oauthor{\bsnm{{Harms}}, \binits{J.}},
\oauthor{\bsnm{{Mapelli}}, \binits{M.}},
\oauthor{\bsnm{{Ronchini}}, \binits{S.}},
\oauthor{\bsnm{{Santoliquido}}, \binits{F.}}:
{Detecting VHE prompt emission from binary neutron-star mergers: ET and CTA
  synergies}.
arXiv e-prints,
2212--14007
(2022)
\doiurl{10.48550/arXiv.2212.14007}
{\href{https://arxiv.org/abs/2212.14007}{{arXiv:2212.14007}}}
{[astro-ph.HE]}
\end{botherref}
\endbibitem

\bibitem[\protect\citeauthoryear{Ronchini et~al.}{2022}]{Ronchini2022}
\begin{barticle}
\bauthor{\bsnm{Ronchini}, \binits{S.}},
\bauthor{\bsnm{Branchesi}, \binits{M.}},
\bauthor{\bsnm{Oganesyan}, \binits{G.}},
\bauthor{\bsnm{Banerjee}, \binits{B.}},
\bauthor{\bsnm{Dupletsa}, \binits{U.}},
\bauthor{\bsnm{Ghirlanda}, \binits{G.}},
\bauthor{\bsnm{Harms}, \binits{J.}},
\bauthor{\bsnm{Mapelli}, \binits{M.}},
\bauthor{\bsnm{Santoliquido}, \binits{F.}}:
\batitle{Perspectives for multimessenger astronomy with the next generation of
  gravitational-wave detectors and high-energy satellites}.
\bjtitle{Astronomy {\&} Astrophysics}
\bvolume{665},
\bfpage{97}
(\byear{2022})
\doiurl{10.1051/0004-6361/202243705}
\end{barticle}
\endbibitem

\bibitem[\protect\citeauthoryear{Iacovelli et~al.}{2022}]{Iacovelli2022}
\begin{barticle}
\bauthor{\bsnm{Iacovelli}, \binits{F.}},
\bauthor{\bsnm{Mancarella}, \binits{M.}},
\bauthor{\bsnm{Foffa}, \binits{S.}},
\bauthor{\bsnm{Maggiore}, \binits{M.}}:
\batitle{Forecasting the detection capabilities of third-generation
  gravitational-wave detectors using {GWFAST}}.
\bjtitle{The Astrophysical Journal}
\bvolume{941}(\bissue{2}),
\bfpage{208}
(\byear{2022})
\doiurl{10.3847/1538-4357/ac9cd4}
\end{barticle}
\endbibitem

\bibitem[\protect\citeauthoryear{{Borhanian} and
  {Sathyaprakash}}{2022}]{Satya2022}
\begin{botherref}
\oauthor{\bsnm{{Borhanian}}, \binits{S.}},
\oauthor{\bsnm{{Sathyaprakash}}, \binits{B.S.}}:
{Listening to the Universe with Next Generation Ground-Based Gravitational-Wave
  Detectors}.
arXiv e-prints,
2202--11048
(2022)
\doiurl{10.48550/arXiv.2202.11048}
{\href{https://arxiv.org/abs/2202.11048}{{arXiv:2202.11048}}}
{[gr-qc]}
\end{botherref}
\endbibitem

\bibitem[\protect\citeauthoryear{Regimbau and Hughes}{2009}]{Regimbau2009}
\begin{botherref}
\oauthor{\bsnm{Regimbau}, \binits{T.}},
\oauthor{\bsnm{Hughes}, \binits{S.A.}}:
Gravitational-wave confusion background from cosmological compact binaries:
  Implications for future terrestrial detectors.
Physical Review D
\textbf{79}(6)
(2009)
\doiurl{10.1103/physrevd.79.062002}
\end{botherref}
\endbibitem

\bibitem[\protect\citeauthoryear{Nitz and Canton}{2021}]{Nitz2021}
\begin{barticle}
\bauthor{\bsnm{Nitz}, \binits{A.H.}},
\bauthor{\bsnm{Canton}, \binits{T.D.}}:
\batitle{Pre-merger localization of compact-binary mergers with
  third-generation observatories}.
\bjtitle{The Astrophysical Journal Letters}
\bvolume{917}(\bissue{2}),
\bfpage{27}
(\byear{2021})
\doiurl{10.3847/2041-8213/ac1a75}
\end{barticle}
\endbibitem

\bibitem[\protect\citeauthoryear{Samajdar et~al.}{2021}]{Samajdar2021}
\begin{botherref}
\oauthor{\bsnm{Samajdar}, \binits{A.}},
\oauthor{\bsnm{Janquart}, \binits{J.}},
\oauthor{\bsnm{Broeck}, \binits{C.V.D.}},
\oauthor{\bsnm{Dietrich}, \binits{T.}}:
Biases in parameter estimation from overlapping gravitational-wave signals in
  the third-generation detector era.
Physical Review D
\textbf{104}(4)
(2021)
\doiurl{10.1103/physrevd.104.044003}
\end{botherref}
\endbibitem

\bibitem[\protect\citeauthoryear{Pizzati et~al.}{2022}]{Pizzati2022}
\begin{botherref}
\oauthor{\bsnm{Pizzati}, \binits{E.}},
\oauthor{\bsnm{Sachdev}, \binits{S.}},
\oauthor{\bsnm{Gupta}, \binits{A.}},
\oauthor{\bsnm{Sathyaprakash}, \binits{B.S.}}:
Toward inference of overlapping gravitational-wave signals.
Physical Review D
\textbf{105}(10)
(2022)
\doiurl{10.1103/physrevd.105.104016}
\end{botherref}
\endbibitem

\bibitem[\protect\citeauthoryear{Himemoto et~al.}{2021}]{Himemoto2021}
\begin{botherref}
\oauthor{\bsnm{Himemoto}, \binits{Y.}},
\oauthor{\bsnm{Nishizawa}, \binits{A.}},
\oauthor{\bsnm{Taruya}, \binits{A.}}:
Impacts of overlapping gravitational-wave signals on the parameter estimation:
  Toward the search for cosmological backgrounds.
Physical Review D
\textbf{104}(4)
(2021)
\doiurl{10.1103/physrevd.104.044010}
\end{botherref}
\endbibitem

\bibitem[\protect\citeauthoryear{Antonelli et~al.}{2021}]{Antonelli2021}
\begin{barticle}
\bauthor{\bsnm{Antonelli}, \binits{A.}},
\bauthor{\bsnm{Burke}, \binits{O.}},
\bauthor{\bsnm{Gair}, \binits{J.R.}}:
\batitle{Noisy neighbours: inference biases from overlapping gravitational-wave
  signals}.
\bjtitle{Monthly Notices of the Royal Astronomical Society}
\bvolume{507}(\bissue{4}),
\bfpage{5069}--\blpage{5086}
(\byear{2021})
\doiurl{10.1093/mnras/stab2358}
\end{barticle}
\endbibitem

\bibitem[\protect\citeauthoryear{Relton and Raymond}{2021}]{Relton2021}
\begin{botherref}
\oauthor{\bsnm{Relton}, \binits{P.}},
\oauthor{\bsnm{Raymond}, \binits{V.}}:
Parameter estimation bias from overlapping binary black hole events in second
  generation interferometers.
Physical Review D
\textbf{104}(8)
(2021)
\doiurl{10.1103/physrevd.104.084039}
\end{botherref}
\endbibitem

\bibitem[\protect\citeauthoryear{Relton et~al.}{2022}]{Relton2022}
\begin{botherref}
\oauthor{\bsnm{Relton}, \binits{P.}},
\oauthor{\bsnm{Virtuoso}, \binits{A.}},
\oauthor{\bsnm{Bini}, \binits{S.}},
\oauthor{\bsnm{Raymond}, \binits{V.}},
\oauthor{\bsnm{Harry}, \binits{I.}},
\oauthor{\bsnm{Drago}, \binits{M.}},
\oauthor{\bsnm{Lazzaro}, \binits{C.}},
\oauthor{\bsnm{Miani}, \binits{A.}},
\oauthor{\bsnm{Tiwari}, \binits{S.}}:
Addressing the challenges of detecting time-overlapping compact binary
  coalescences.
Physical Review D
\textbf{106}(10)
(2022)
\doiurl{10.1103/physrevd.106.104045}
\end{botherref}
\endbibitem

\bibitem[\protect\citeauthoryear{Collaboration}{2022}]{Virgo_nEXT}
\begin{barticle}
\bauthor{\bsnm{Collaboration}, \binits{T.V.}}:
\batitle{Virgo next: beyond the adv+ project a concept study}.
\bjtitle{Internal document}
(\byear{2022})
\doiurl{VIR-0497A-22}
\end{barticle}
\endbibitem

\bibitem[\protect\citeauthoryear{Adhikari et~al.}{2020}]{LIGO_Voyager}
\begin{barticle}
\bauthor{\bsnm{Adhikari}, \binits{R.X.}}, \betal:
\batitle{A cryogenic silicon interferometer for gravitational-wave detection}.
\bjtitle{Classical and Quantum Gravity}
\bvolume{37}(\bissue{16}),
\bfpage{165003}
(\byear{2020})
\doiurl{10.1088/1361-6382/ab9143}
\end{barticle}
\endbibitem

\bibitem[\protect\citeauthoryear{Comon and Jutten}{2010}]{comon2010handbook}
\begin{bbook}
\bauthor{\bsnm{Comon}, \binits{P.}},
\bauthor{\bsnm{Jutten}, \binits{C.}}:
\bbtitle{Handbook of Blind Source Separation: Independent Component Analysis
  and Applications}.
\bpublisher{Academic press}, \blocation{???}
(\byear{2010})
\end{bbook}
\endbibitem

\bibitem[\protect\citeauthoryear{Pan et~al.}{2021}]{Pan2021}
\begin{botherref}
\oauthor{\bsnm{Pan}, \binits{Y.}},
\oauthor{\bsnm{Matilainen}, \binits{M.}},
\oauthor{\bsnm{Taskinen}, \binits{S.}},
\oauthor{\bsnm{Nordhausen}, \binits{K.}}:
A review of second-order blind identification methods.
{WIREs} Computational Statistics
\textbf{14}(4)
(2021)
\doiurl{10.1002/wics.1550}
\end{botherref}
\endbibitem

\bibitem[\protect\citeauthoryear{Aapo~Hyv\"{a}rinen}{2001}]{Oja2001}
\begin{bbook}
\bauthor{\bsnm{Aapo~Hyv\"{a}rinen}, \binits{E.O.} \bsuffix{Juha~Karhunen}}:
\bbtitle{Independent Component Analysis}.
\bpublisher{John Wiley \& sons}, \blocation{???}
(\byear{2001})
\end{bbook}
\endbibitem

\bibitem[\protect\citeauthoryear{Nordhausen and Oja}{2018}]{Nordhausen2018}
\begin{botherref}
\oauthor{\bsnm{Nordhausen}, \binits{K.}},
\oauthor{\bsnm{Oja}, \binits{H.}}:
Independent component analysis: A statistical perspective.
WIREs Computational Statistics
\textbf{10}(5)
(2018)
\doiurl{10.1002/wics.1440}
\end{botherref}
\endbibitem

\bibitem[\protect\citeauthoryear{Yuan et~al.}{2021}]{Yuan2021}
\begin{barticle}
\bauthor{\bsnm{Yuan}, \binits{H.}},
\bauthor{\bsnm{Wu}, \binits{N.}},
\bauthor{\bsnm{Chen}, \binits{X.}}:
\batitle{Mechanical compound fault analysis method based on shift invariant
  dictionary learning and improved {FastICA} algorithm}.
\bjtitle{Machines}
\bvolume{9}(\bissue{8}),
\bfpage{144}
(\byear{2021})
\doiurl{10.3390/machines9080144}
\end{barticle}
\endbibitem

\bibitem[\protect\citeauthoryear{Uddin et~al.}{2021}]{Uddin2021}
\begin{barticle}
\bauthor{\bsnm{Uddin}, \binits{Z.}},
\bauthor{\bsnm{Qamar}, \binits{A.}},
\bauthor{\bsnm{Alam}, \binits{F.}}:
\batitle{{ICA} based sensors fault diagnosis: An audio separation application}.
\bjtitle{Wireless Personal Communications}
\bvolume{118}(\bissue{4}),
\bfpage{3369}--\blpage{3384}
(\byear{2021})
\doiurl{10.1007/s11277-021-08184-x}
\end{barticle}
\endbibitem

\bibitem[\protect\citeauthoryear{Liao and Niebur}{2003}]{HuaiweiLiao2003}
\begin{barticle}
\bauthor{\bsnm{Liao}, \binits{H.}},
\bauthor{\bsnm{Niebur}, \binits{D.}}:
\batitle{Load profile estimation in electric transmission networks using
  independent component analysis}.
\bjtitle{{IEEE} Transactions on Power Systems}
\bvolume{18}(\bissue{2}),
\bfpage{707}--\blpage{715}
(\byear{2003})
\doiurl{10.1109/tpwrs.2003.811199}
\end{barticle}
\endbibitem

\bibitem[\protect\citeauthoryear{Wu et~al.}{2021}]{wu2021peak}
\begin{barticle}
\bauthor{\bsnm{Wu}, \binits{Y.}},
\bauthor{\bsnm{Wu}, \binits{K.}},
\bauthor{\bsnm{Li}, \binits{W.}},
\bauthor{\bsnm{Chen}, \binits{J.}},
\bauthor{\bsnm{Yu}, \binits{Z.}}:
\batitle{Peak-load-regulation nuclear power unit fault diagnosis using thermal
  sensors combined with improved ica-rf algorithm}.
\bjtitle{Sensors}
\bvolume{21}(\bissue{21}),
\bfpage{6955}
(\byear{2021})
\end{barticle}
\endbibitem

\bibitem[\protect\citeauthoryear{Kim et~al.}{2022}]{Kim2022}
\begin{barticle}
\bauthor{\bsnm{Kim}, \binits{M.}},
\bauthor{\bsnm{Lee}, \binits{S.-H.}},
\bauthor{\bsnm{Choi}, \binits{I.-O.}},
\bauthor{\bsnm{Kim}, \binits{K.-T.}}:
\batitle{Direction finding for multiple wideband chirp signal sources using
  blind signal separation and matched filtering}.
\bjtitle{Signal Processing}
\bvolume{200},
\bfpage{108642}
(\byear{2022})
\doiurl{10.1016/j.sigpro.2022.108642}
\end{barticle}
\endbibitem

\bibitem[\protect\citeauthoryear{Bharadwaj
  et~al.}{2017}]{bharadwaj2017deblending}
\begin{bchapter}
\bauthor{\bsnm{Bharadwaj}, \binits{P.}},
\bauthor{\bsnm{Demanet}, \binits{L.}},
\bauthor{\bsnm{Fournier}, \binits{A.}}:
\bctitle{Deblending random seismic sources via independent component analysis}.
In: \bbtitle{SEG Technical Program Expanded Abstracts 2017},
pp. \bfpage{4898}--\blpage{4902}.
\bpublisher{Society of Exploration Geophysicists}, \blocation{???}
(\byear{2017})
\end{bchapter}
\endbibitem

\bibitem[\protect\citeauthoryear{Lubo-Robles and
  Marfurt}{2019}]{lubo2019independent}
\begin{barticle}
\bauthor{\bsnm{Lubo-Robles}, \binits{D.}},
\bauthor{\bsnm{Marfurt}, \binits{K.J.}}:
\batitle{Independent component analysis for reservoir geomorphology and
  unsupervised seismic facies classification in the taranaki basin, new
  zealand}.
\bjtitle{Interpretation}
\bvolume{7}(\bissue{3}),
\bfpage{19}--\blpage{42}
(\byear{2019})
\end{barticle}
\endbibitem

\bibitem[\protect\citeauthoryear{James and Hesse}{2004}]{james2004independent}
\begin{barticle}
\bauthor{\bsnm{James}, \binits{C.J.}},
\bauthor{\bsnm{Hesse}, \binits{C.W.}}:
\batitle{Independent component analysis for biomedical signals}.
\bjtitle{Physiological measurement}
\bvolume{26}(\bissue{1}),
\bfpage{15}
(\byear{2004})
\end{barticle}
\endbibitem

\bibitem[\protect\citeauthoryear{Hsieh et~al.}{2009}]{hsieh2009independent}
\begin{bchapter}
\bauthor{\bsnm{Hsieh}, \binits{H.-L.}},
\bauthor{\bsnm{Chien}, \binits{J.-T.}},
\bauthor{\bsnm{Shinoda}, \binits{K.}},
\bauthor{\bsnm{Furui}, \binits{S.}}:
\bctitle{Independent component analysis for noisy speech recognition}.
In: \bbtitle{2009 IEEE International Conference on Acoustics, Speech and Signal
  Processing},
pp. \bfpage{4369}--\blpage{4372}
(\byear{2009}).
\bcomment{IEEE}
\end{bchapter}
\endbibitem

\bibitem[\protect\citeauthoryear{Uddin et~al.}{2015}]{uddin2015applications}
\begin{barticle}
\bauthor{\bsnm{Uddin}, \binits{Z.}},
\bauthor{\bsnm{Ahmad}, \binits{A.}},
\bauthor{\bsnm{Iqbal}, \binits{M.}},
\bauthor{\bsnm{Naeem}, \binits{M.}}:
\batitle{Applications of independent component analysis in wireless
  communication systems}.
\bjtitle{Wireless personal communications}
\bvolume{83}(\bissue{4}),
\bfpage{2711}--\blpage{2737}
(\byear{2015})
\end{barticle}
\endbibitem

\bibitem[\protect\citeauthoryear{De~Rosa et~al.}{2012}]{derosa2012improvement}
\begin{barticle}
\bauthor{\bsnm{De~Rosa}, \binits{R.}},
\bauthor{\bsnm{Forte}, \binits{L.}},
\bauthor{\bsnm{Garufi}, \binits{F.}},
\bauthor{\bsnm{Milano}, \binits{L.}}:
\batitle{Improvement of the performance of a classical matched filter by an
  independent component analysis preprocessing}.
\bjtitle{Physical Review D}
\bvolume{85}(\bissue{4}),
\bfpage{042001}
(\byear{2012})
\end{barticle}
\endbibitem

\bibitem[\protect\citeauthoryear{Morisaki et~al.}{2016}]{morisaki2016toward}
\begin{barticle}
\bauthor{\bsnm{Morisaki}, \binits{S.}},
\bauthor{\bsnm{Yokoyama}, \binits{J.}},
\bauthor{\bsnm{Eda}, \binits{K.}},
\bauthor{\bsnm{Itoh}, \binits{Y.}}:
\batitle{Toward the detection of gravitational waves under non-gaussian noises
  ii. independent component analysis}.
\bjtitle{Proceedings of the Japan Academy, Series B}
\bvolume{92}(\bissue{8}),
\bfpage{336}--\blpage{345}
(\byear{2016})
\end{barticle}
\endbibitem

\bibitem[\protect\citeauthoryear{Akutsu et~al.}{2020}]{akutsu2020application}
\begin{barticle}
\bauthor{\bsnm{Akutsu}, \binits{T.}},
\bauthor{\bsnm{Ando}, \binits{M.}},
\bauthor{\bsnm{Arai}, \binits{K.}},
\bauthor{\bsnm{Arai}, \binits{Y.}},
\bauthor{\bsnm{Araki}, \binits{S.}},
\bauthor{\bsnm{Araya}, \binits{A.}},
\bauthor{\bsnm{Aritomi}, \binits{N.}},
\bauthor{\bsnm{Asada}, \binits{H.}},
\bauthor{\bsnm{Aso}, \binits{Y.}},
\bauthor{\bsnm{Atsuta}, \binits{S.}}, \betal:
\batitle{Application of independent component analysis to the ikagra data}.
\bjtitle{Progress of Theoretical and Experimental Physics}
\bvolume{2020}(\bissue{5}),
\bfpage{053}--\blpage{01}
(\byear{2020})
\end{barticle}
\endbibitem

\bibitem[\protect\citeauthoryear{Liebisch et~al.}{2021}]{Liebisch2021}
\begin{barticle}
\bauthor{\bsnm{Liebisch}, \binits{A.P.}},
\bauthor{\bsnm{Eggert}, \binits{T.}},
\bauthor{\bsnm{Shindy}, \binits{A.}},
\bauthor{\bsnm{Valentini}, \binits{E.}},
\bauthor{\bsnm{Irving}, \binits{S.}},
\bauthor{\bsnm{Stankewitz}, \binits{A.}},
\bauthor{\bsnm{Schulz}, \binits{E.}}:
\batitle{A novel tool for the removal of muscle artefacts from {EEG}: Improving
  data quality in the gamma frequency range}.
\bjtitle{Journal of Neuroscience Methods}
\bvolume{358},
\bfpage{109217}
(\byear{2021})
\doiurl{10.1016/j.jneumeth.2021.109217}
\end{barticle}
\endbibitem

\bibitem[\protect\citeauthoryear{Wu et~al.}{2021}]{Wu2021}
\begin{barticle}
\bauthor{\bsnm{Wu}, \binits{B.}},
\bauthor{\bsnm{Pal}, \binits{S.}},
\bauthor{\bsnm{Kang}, \binits{J.}},
\bauthor{\bsnm{Guo}, \binits{Y.}}:
\batitle{Distributional independent component analysis for diverse neuroimaging
  modalities}.
\bjtitle{Biometrics}
\bvolume{78}(\bissue{3}),
\bfpage{1092}--\blpage{1105}
(\byear{2021})
\doiurl{10.1111/biom.13594}
\end{barticle}
\endbibitem

\bibitem[\protect\citeauthoryear{Kim and Yoo}{2004}]{Kim2004underdetrmined}
\begin{bchapter}
\bauthor{\bsnm{Kim}, \binits{S.G.}},
\bauthor{\bsnm{Yoo}, \binits{C.D.}}:
\bctitle{Underdetermined independent component analysis by data generation}.
In: \bbtitle{Independent Component Analysis and Blind Signal Separation},
pp. \bfpage{445}--\blpage{452}.
\bpublisher{Springer}, \blocation{???}
(\byear{2004}).
\doiurl{10.1007/978-3-540-30110-3_57} .
\burl{https://doi.org/10.1007/978-3-540-30110-3_57}
\end{bchapter}
\endbibitem

\bibitem[\protect\citeauthoryear{Stone}{2004}]{stone2004}
\begin{bbook}
\bauthor{\bsnm{Stone}, \binits{J.V.}}:
\bbtitle{Independent Component Analysis: a Tutorial Introduction}.
\bpublisher{MIT press}, \blocation{???}
(\byear{2004})
\end{bbook}
\endbibitem

\bibitem[\protect\citeauthoryear{Hyv{\"a}rinen and
  Oja}{2000}]{hyvarinen2000independent}
\begin{barticle}
\bauthor{\bsnm{Hyv{\"a}rinen}, \binits{A.}},
\bauthor{\bsnm{Oja}, \binits{E.}}:
\batitle{Independent component analysis: algorithms and applications}.
\bjtitle{Neural networks}
\bvolume{13}(\bissue{4-5}),
\bfpage{411}--\blpage{430}
(\byear{2000})
\end{barticle}
\endbibitem

\bibitem[\protect\citeauthoryear{Cardoso}{1997}]{cardoso1997infomax}
\begin{barticle}
\bauthor{\bsnm{Cardoso}, \binits{J.-F.}}:
\batitle{Infomax and maximum likelihood for blind source separation}.
\bjtitle{IEEE Signal processing letters}
\bvolume{4}(\bissue{4}),
\bfpage{112}--\blpage{114}
(\byear{1997})
\end{barticle}
\endbibitem

\bibitem[\protect\citeauthoryear{Hyvarinen}{1999}]{Hyvarinen1999}
\begin{barticle}
\bauthor{\bsnm{Hyvarinen}, \binits{A.}}:
\batitle{Fast and robust fixed-point algorithms for independent component
  analysis}.
\bjtitle{{IEEE} Transactions on Neural Networks}
\bvolume{10}(\bissue{3}),
\bfpage{626}--\blpage{634}
(\byear{1999})
\doiurl{10.1109/72.761722}
\end{barticle}
\endbibitem

\bibitem[\protect\citeauthoryear{Bell and Sejnowski}{1995}]{Bell1995}
\begin{barticle}
\bauthor{\bsnm{Bell}, \binits{A.J.}},
\bauthor{\bsnm{Sejnowski}, \binits{T.J.}}:
\batitle{An information-maximization approach to blind separation and blind
  deconvolution}.
\bjtitle{Neural Computation}
\bvolume{7}(\bissue{6}),
\bfpage{1129}--\blpage{1159}
(\byear{1995})
\doiurl{10.1162/neco.1995.7.6.1129}
\end{barticle}
\endbibitem

\bibitem[\protect\citeauthoryear{Cardoso and Souloumiac}{1993}]{Cardoso1993}
\begin{barticle}
\bauthor{\bsnm{Cardoso}, \binits{J.F.}},
\bauthor{\bsnm{Souloumiac}, \binits{A.}}:
\batitle{Blind beamforming for non-gaussian signals}.
\bjtitle{{IEE} Proceedings F Radar and Signal Processing}
\bvolume{140}(\bissue{6}),
\bfpage{362}
(\byear{1993})
\doiurl{10.1049/ip-f-2.1993.0054}
\end{barticle}
\endbibitem

\bibitem[\protect\citeauthoryear{Hyvärinen}{1998}]{HYVARINEN199849}
\begin{barticle}
\bauthor{\bsnm{Hyvärinen}, \binits{A.}}:
\batitle{Independent component analysis in the presence of gaussian noise by
  maximizing joint likelihood}.
\bjtitle{Neurocomputing}
\bvolume{22}(\bissue{1}),
\bfpage{49}--\blpage{67}
(\byear{1998})
\doiurl{10.1016/S0925-2312(98)00049-6}
\end{barticle}
\endbibitem

\bibitem[\protect\citeauthoryear{Hyvarinen}{1999}]{hyvarinen1999noisyGaussian}
\begin{bchapter}
\bauthor{\bsnm{Hyvarinen}, \binits{A.}}:
\bctitle{Fast {ICA} for noisy data using gaussian moments}.
In: \bbtitle{1999 IEEE International Symposium on Circuits and Systems
  (ISCAS)},
vol. \bseriesno{5},
pp. \bfpage{57}--\blpage{61}
(\byear{1999}).
\bcomment{IEEE}
\end{bchapter}
\endbibitem

\bibitem[\protect\citeauthoryear{Cao et~al.}{2003}]{Cao_2003}
\begin{barticle}
\bauthor{\bsnm{Cao}, \binits{J.}},
\bauthor{\bsnm{Murata}, \binits{N.}},
\bauthor{\bsnm{Amari}, \binits{S.}},
\bauthor{\bsnm{Cichocki}, \binits{A.}},
\bauthor{\bsnm{Takeda}, \binits{T.}}:
\batitle{A robust approach to independent component analysis of signals with
  high-level noise measurements}.
\bjtitle{IEEE Transactions on Neural Networks}
\bvolume{14}(\bissue{3}),
\bfpage{631}--\blpage{645}
(\byear{2003})
\doiurl{10.1109/TNN.2002.806648}
\end{barticle}
\endbibitem

\bibitem[\protect\citeauthoryear{Tharwat}{2020}]{tharwat2020ICA}
\begin{botherref}
\oauthor{\bsnm{Tharwat}, \binits{A.}}:
Independent component analysis: An introduction.
Applied Computing and Informatics
(2020)
\end{botherref}
\endbibitem

\bibitem[\protect\citeauthoryear{Soroush et~al.}{2022}]{ZangenehSoroush2022}
\begin{botherref}
\oauthor{\bsnm{Soroush}, \binits{M.Z.}},
\oauthor{\bsnm{Tahvilian}, \binits{P.}},
\oauthor{\bsnm{Nasirpour}, \binits{M.H.}},
\oauthor{\bsnm{Maghooli}, \binits{K.}},
\oauthor{\bsnm{Sadeghniiat-Haghighi}, \binits{K.}},
\oauthor{\bsnm{Harandi}, \binits{S.V.}},
\oauthor{\bsnm{Abdollahi}, \binits{Z.}},
\oauthor{\bsnm{Ghazizadeh}, \binits{A.}},
\oauthor{\bsnm{Dabanloo}, \binits{N.J.}}:
{EEG} artifact removal using sub-space decomposition, nonlinear dynamics,
  stationary wavelet transform and machine learning algorithms.
Frontiers in Physiology
\textbf{13}
(2022)
\doiurl{10.3389/fphys.2022.910368}
\end{botherref}
\endbibitem

\bibitem[\protect\citeauthoryear{Wu et~al.}{2021}]{HWu2021}
\begin{barticle}
\bauthor{\bsnm{Wu}, \binits{H.}},
\bauthor{\bsnm{Li}, \binits{D.}},
\bauthor{\bsnm{Lu}, \binits{M.}},
\bauthor{\bsnm{Zeng}, \binits{Y.}}:
\batitle{{fMRI} activations via low-complexity second-order
  inverse-sparse-transform blind separation}.
\bjtitle{Digital Signal Processing}
\bvolume{117},
\bfpage{103137}
(\byear{2021})
\doiurl{10.1016/j.dsp.2021.103137}
\end{barticle}
\endbibitem

\bibitem[\protect\citeauthoryear{Lakshmi et~al.}{2021}]{Lakshmi2021}
\begin{barticle}
\bauthor{\bsnm{Lakshmi}, \binits{K.}},
\bauthor{\bsnm{Reddy}, \binits{V.K.}},
\bauthor{\bsnm{Rao}, \binits{A.R.M.}}:
\batitle{Modal identification of practical engineering structures using
  second-order blind identification}.
\bjtitle{Journal of The Institution of Engineers (India): Series A}
\bvolume{102}(\bissue{2}),
\bfpage{499}--\blpage{512}
(\byear{2021})
\doiurl{10.1007/s40030-021-00523-2}
\end{barticle}
\endbibitem

\bibitem[\protect\citeauthoryear{Eshkevari et~al.}{2020}]{SadeghiEshkevari2020}
\begin{barticle}
\bauthor{\bsnm{Eshkevari}, \binits{S.S.}},
\bauthor{\bsnm{Matarazzo}, \binits{T.J.}},
\bauthor{\bsnm{Pakzad}, \binits{S.N.}}:
\batitle{Bridge modal identification using acceleration measurements within
  moving vehicles}.
\bjtitle{Mechanical Systems and Signal Processing}
\bvolume{141},
\bfpage{106733}
(\byear{2020})
\doiurl{10.1016/j.ymssp.2020.106733}
\end{barticle}
\endbibitem

\bibitem[\protect\citeauthoryear{de~Oliveira et~al.}{2021}]{deOliveira2021}
\begin{barticle}
\bauthor{\bsnm{Oliveira}, \binits{D.R.}},
\bauthor{\bsnm{Lima}, \binits{M.A.A.}},
\bauthor{\bsnm{Silva}, \binits{L.R.M.}},
\bauthor{\bsnm{Ferreira}, \binits{D.D.}},
\bauthor{\bsnm{Duque}, \binits{C.A.}}:
\batitle{Second order blind identification algorithm with exact model order
  estimation for harmonic and interharmonic decomposition with reduced
  complexity}.
\bjtitle{International Journal of Electrical Power and Energy Systems}
\bvolume{125},
\bfpage{106415}
(\byear{2021})
\doiurl{10.1016/j.ijepes.2020.106415}
\end{barticle}
\endbibitem

\bibitem[\protect\citeauthoryear{Belouchrani et~al.}{1997}]{Belouchrani1997}
\begin{barticle}
\bauthor{\bsnm{Belouchrani}, \binits{A.}},
\bauthor{\bsnm{Abed-Meraim}, \binits{K.}},
\bauthor{\bsnm{Cardoso}, \binits{J.-F.}},
\bauthor{\bsnm{Moulines}, \binits{E.}}:
\batitle{A blind source separation technique using second-order statistics}.
\bjtitle{{IEEE} Transactions on Signal Processing}
\bvolume{45}(\bissue{2}),
\bfpage{434}--\blpage{444}
(\byear{1997})
\doiurl{10.1109/78.554307}
\end{barticle}
\endbibitem

\bibitem[\protect\citeauthoryear{Tong et~al.}{}]{Tong}
\begin{botherref}
\oauthor{\bsnm{Tong}, \binits{L.}},
\oauthor{\bsnm{Soon}, \binits{V.C.}},
\oauthor{\bsnm{Huang}, \binits{Y.F.}},
\oauthor{\bsnm{Liu}, \binits{R.}}:
{AMUSE}: a new blind identification algorithm.
In: {IEEE} International Symposium on Circuits and Systems.
{IEEE}.
\doiurl{10.1109/iscas.1990.111981} .
\url{https://doi.org/10.1109/iscas.1990.111981}
\end{botherref}
\endbibitem

\bibitem[\protect\citeauthoryear{Wang et~al.}{2023}]{Wang2023}
\begin{barticle}
\bauthor{\bsnm{Wang}, \binits{W.}},
\bauthor{\bsnm{Wang}, \binits{S.}},
\bauthor{\bsnm{Qin}, \binits{D.}},
\bauthor{\bsnm{Fang}, \binits{Y.}},
\bauthor{\bsnm{Zheng}, \binits{Y.}}:
\batitle{Heart-lung sound separation by nonnegative matrix factorization and
  deep learning}.
\bjtitle{Biomedical Signal Processing and Control}
\bvolume{79},
\bfpage{104180}
(\byear{2023})
\doiurl{10.1016/j.bspc.2022.104180}
\end{barticle}
\endbibitem

\bibitem[\protect\citeauthoryear{Leplat et~al.}{2022}]{Leplat2022}
\begin{barticle}
\bauthor{\bsnm{Leplat}, \binits{V.}},
\bauthor{\bsnm{Gillis}, \binits{N.}},
\bauthor{\bsnm{F{\'{e}}votte}, \binits{C.}}:
\batitle{Multi-resolution beta-divergence {NMF} for blind spectral unmixing}.
\bjtitle{Signal Processing}
\bvolume{193},
\bfpage{108428}
(\byear{2022})
\doiurl{10.1016/j.sigpro.2021.108428}
\end{barticle}
\endbibitem

\bibitem[\protect\citeauthoryear{Yakimov et~al.}{2021}]{Yakimov2021}
\begin{barticle}
\bauthor{\bsnm{Yakimov}, \binits{B.P.}},
\bauthor{\bsnm{Venets}, \binits{A.V.}},
\bauthor{\bsnm{Schleusener}, \binits{J.}},
\bauthor{\bsnm{Fadeev}, \binits{V.V.}},
\bauthor{\bsnm{Lademann}, \binits{J.}},
\bauthor{\bsnm{Shirshin}, \binits{E.A.}},
\bauthor{\bsnm{Darvin}, \binits{M.E.}}:
\batitle{Blind source separation of molecular components of the human skin in
  vivo: non-negative matrix factorization of raman microspectroscopy data}.
\bjtitle{The Analyst}
\bvolume{146}(\bissue{10}),
\bfpage{3185}--\blpage{3196}
(\byear{2021})
\doiurl{10.1039/d0an02480e}
\end{barticle}
\endbibitem

\bibitem[\protect\citeauthoryear{Gu et~al.}{2021}]{Gu2021}
\begin{barticle}
\bauthor{\bsnm{Gu}, \binits{M.}},
\bauthor{\bsnm{Xie}, \binits{R.}},
\bauthor{\bsnm{Jin}, \binits{G.}},
\bauthor{\bsnm{Xu}, \binits{C.}},
\bauthor{\bsnm{Wang}, \binits{S.}},
\bauthor{\bsnm{Liu}, \binits{J.}},
\bauthor{\bsnm{Wei}, \binits{H.}}:
\batitle{Quantitative evaluation for fluid components on 2d {NMR} spectrum
  using blind source separation}.
\bjtitle{Journal of Magnetic Resonance}
\bvolume{332},
\bfpage{107079}
(\byear{2021})
\doiurl{10.1016/j.jmr.2021.107079}
\end{barticle}
\endbibitem

\bibitem[\protect\citeauthoryear{Wei et~al.}{2021}]{Wei2021}
\begin{barticle}
\bauthor{\bsnm{Wei}, \binits{J.}},
\bauthor{\bsnm{Chen}, \binits{P.}},
\bauthor{\bsnm{Han}, \binits{Y.}},
\bauthor{\bsnm{Zhao}, \binits{Y.}}:
\batitle{Blind separation model of multi-voltage projections for the hardening
  artifact correction in computed tomography}.
\bjtitle{Biomedical Signal Processing and Control}
\bvolume{64},
\bfpage{102236}
(\byear{2021})
\doiurl{10.1016/j.bspc.2020.102236}
\end{barticle}
\endbibitem

\bibitem[\protect\citeauthoryear{Flamant et~al.}{2020}]{Flamant2020}
\begin{barticle}
\bauthor{\bsnm{Flamant}, \binits{J.}},
\bauthor{\bsnm{Miron}, \binits{S.}},
\bauthor{\bsnm{Brie}, \binits{D.}}:
\batitle{Quaternion non-negative matrix factorization: Definition, uniqueness,
  and algorithm}.
\bjtitle{{IEEE} Transactions on Signal Processing}
\bvolume{68},
\bfpage{1870}--\blpage{1883}
(\byear{2020})
\doiurl{10.1109/tsp.2020.2974651}
\end{barticle}
\endbibitem

\bibitem[\protect\citeauthoryear{Hern{\'{a}}ndez-Villegas
  et~al.}{2019}]{HernndezVillegas2019}
\begin{botherref}
\oauthor{\bsnm{Hern{\'{a}}ndez-Villegas}, \binits{Y.}},
\oauthor{\bsnm{Ortega-Martorell}, \binits{S.}},
\oauthor{\bsnm{Ar{\'{u}}s}, \binits{C.}},
\oauthor{\bsnm{Vellido}, \binits{A.}},
\oauthor{\bsnm{Juli{\`{a}}-Sap{\'{e}}}, \binits{M.}}:
Extraction of artefactual {MRS} patterns from a large database using
  non-negative matrix factorization.
{NMR} in Biomedicine
\textbf{35}(4)
(2019)
\doiurl{10.1002/nbm.4193}
\end{botherref}
\endbibitem

\bibitem[\protect\citeauthoryear{Song et~al.}{2020}]{Song2020}
\begin{barticle}
\bauthor{\bsnm{Song}, \binits{J.E.}},
\bauthor{\bsnm{Shin}, \binits{J.}},
\bauthor{\bsnm{Lee}, \binits{H.}},
\bauthor{\bsnm{Lee}, \binits{H.J.}},
\bauthor{\bsnm{Moon}, \binits{W.-J.}},
\bauthor{\bsnm{Kim}, \binits{D.-H.}}:
\batitle{Blind source separation for myelin water fraction mapping using
  multi-echo gradient echo imaging}.
\bjtitle{{IEEE} Transactions on Medical Imaging}
\bvolume{39}(\bissue{6}),
\bfpage{2235}--\blpage{2245}
(\byear{2020})
\doiurl{10.1109/tmi.2020.2967068}
\end{barticle}
\endbibitem

\bibitem[\protect\citeauthoryear{Lyu et~al.}{2020}]{Lyu2020}
\begin{botherref}
\oauthor{\bsnm{Lyu}, \binits{S.}},
\oauthor{\bsnm{Liu}, \binits{Y.}},
\oauthor{\bsnm{Hou}, \binits{M.}},
\oauthor{\bsnm{Yin}, \binits{Q.}},
\oauthor{\bsnm{Wu}, \binits{W.}},
\oauthor{\bsnm{Yang}, \binits{X.}}:
Quantitative analysis of mixed pigments for chinese paintings using the
  improved method of ratio spectra derivative spectrophotometry based on mode.
Heritage Science
\textbf{8}(1)
(2020)
\doiurl{10.1186/s40494-020-00372-5}
\end{botherref}
\endbibitem

\bibitem[\protect\citeauthoryear{Gurve and Krishnan}{2020}]{Gurve2020}
\begin{barticle}
\bauthor{\bsnm{Gurve}, \binits{D.}},
\bauthor{\bsnm{Krishnan}, \binits{S.}}:
\batitle{Separation of fetal-{ECG} from single-channel abdominal {ECG} using
  activation scaled non-negative matrix factorization}.
\bjtitle{{IEEE} Journal of Biomedical and Health Informatics}
\bvolume{24}(\bissue{3}),
\bfpage{669}--\blpage{680}
(\byear{2020})
\doiurl{10.1109/jbhi.2019.2920356}
\end{barticle}
\endbibitem

\bibitem[\protect\citeauthoryear{He et~al.}{2016}]{He2016}
\begin{barticle}
\bauthor{\bsnm{He}, \binits{P.}},
\bauthor{\bsnm{Chen}, \binits{X.}},
\bauthor{\bsnm{Zeng}, \binits{H.}}:
\batitle{Wireless sensor network for multi-target detection algorithm based on
  blind source separation}.
\bjtitle{International Journal of Security and Networks}
\bvolume{11}(\bissue{4}),
\bfpage{235}
(\byear{2016})
\doiurl{10.1504/ijsn.2016.079275}
\end{barticle}
\endbibitem

\bibitem[\protect\citeauthoryear{Li}{2012}]{Li2012}
\begin{barticle}
\bauthor{\bsnm{Li}, \binits{H.}}:
\batitle{Non-negative matrix factorization of mixed speech signals based on
  improved particle swarm optimization}.
\bjtitle{The Journal of the Acoustical Society of America}
\bvolume{131}(\bissue{4}),
\bfpage{3237}--\blpage{3237}
(\byear{2012})
\doiurl{10.1121/1.4708078}
\end{barticle}
\endbibitem

\bibitem[\protect\citeauthoryear{Laudadio et~al.}{2016}]{Laudadio2016}
\begin{barticle}
\bauthor{\bsnm{Laudadio}, \binits{T.}},
\bauthor{\bsnm{Sava}, \binits{A.R.C.}},
\bauthor{\bsnm{Sima}, \binits{D.M.}},
\bauthor{\bsnm{Wright}, \binits{A.J.}},
\bauthor{\bsnm{Heerschap}, \binits{A.}},
\bauthor{\bsnm{Mastronardi}, \binits{N.}},
\bauthor{\bsnm{Huffel}, \binits{S.V.}}:
\batitle{Hierarchical non-negative matrix factorization applied to
  three-dimensional 3 t {MRSI} data for automatic tissue characterization of
  the prostate}.
\bjtitle{{NMR} in Biomedicine}
\bvolume{29}(\bissue{6}),
\bfpage{751}--\blpage{758}
(\byear{2016})
\doiurl{10.1002/nbm.3527}
\end{barticle}
\endbibitem

\bibitem[\protect\citeauthoryear{Gillis}{2020}]{gillis2020nonnegative}
\begin{bbook}
\bauthor{\bsnm{Gillis}, \binits{N.}}:
\bbtitle{Nonnegative Matrix Factorization}.
\bpublisher{SIAM}, \blocation{???}
(\byear{2020})
\end{bbook}
\endbibitem

\bibitem[\protect\citeauthoryear{Lee and Seung}{1999}]{lee1999NMF}
\begin{barticle}
\bauthor{\bsnm{Lee}, \binits{D.D.}},
\bauthor{\bsnm{Seung}, \binits{H.S.}}:
\batitle{Learning the parts of objects by non-negative matrix factorization}.
\bjtitle{Nature}
\bvolume{401}(\bissue{6755}),
\bfpage{788}--\blpage{791}
(\byear{1999})
\end{barticle}
\endbibitem

\bibitem[\protect\citeauthoryear{Vial et~al.}{2021}]{Phase_rec_2021}
\begin{barticle}
\bauthor{\bsnm{Vial}, \binits{P.-H.}},
\bauthor{\bsnm{Magron}, \binits{P.}},
\bauthor{\bsnm{Oberlin}, \binits{T.}},
\bauthor{\bsnm{Févotte}, \binits{C.}}:
\batitle{Phase retrieval with bregman divergences and application to audio
  signal recovery}.
\bjtitle{IEEE Journal of Selected Topics in Signal Processing}
\bvolume{15}(\bissue{1}),
\bfpage{51}--\blpage{64}
(\byear{2021})
\doiurl{10.1109/JSTSP.2021.3051870}
\end{barticle}
\endbibitem

\bibitem[\protect\citeauthoryear{O'grady and
  Pearlmutter}{2006}]{o2006convolutive}
\begin{bchapter}
\bauthor{\bsnm{O'grady}, \binits{P.D.}},
\bauthor{\bsnm{Pearlmutter}, \binits{B.A.}}:
\bctitle{Convolutive non-negative matrix factorisation with a sparseness
  constraint}.
In: \bbtitle{2006 16th IEEE Signal Processing Society Workshop on Machine
  Learning for Signal Processing},
pp. \bfpage{427}--\blpage{432}
(\byear{2006}).
\bcomment{IEEE}
\end{bchapter}
\endbibitem

\bibitem[\protect\citeauthoryear{Yilmaz and Rickard}{2004}]{Yilmaz2004}
\begin{barticle}
\bauthor{\bsnm{Yilmaz}, \binits{O.}},
\bauthor{\bsnm{Rickard}, \binits{S.}}:
\batitle{Blind separation of speech mixtures via time-frequency masking}.
\bjtitle{{IEEE} Transactions on Signal Processing}
\bvolume{52}(\bissue{7}),
\bfpage{1830}--\blpage{1847}
(\byear{2004})
\doiurl{10.1109/tsp.2004.828896}
\end{barticle}
\endbibitem

\bibitem[\protect\citeauthoryear{Yu et~al.}{2016}]{Yu2016}
\begin{botherref}
\oauthor{\bsnm{Yu}, \binits{Y.}},
\oauthor{\bsnm{Wang}, \binits{W.}},
\oauthor{\bsnm{Han}, \binits{P.}}:
Localization based stereo speech source separation using probabilistic
  time-frequency masking and deep neural networks.
{EURASIP} Journal on Audio, Speech, and Music Processing
\textbf{2016}(1)
(2016)
\doiurl{10.1186/s13636-016-0085-x}
\end{botherref}
\endbibitem

\bibitem[\protect\citeauthoryear{Mustafi and Ghorai}{2013}]{Mustafi2013}
\begin{barticle}
\bauthor{\bsnm{Mustafi}, \binits{A.}},
\bauthor{\bsnm{Ghorai}, \binits{S.K.}}:
\batitle{A novel blind source separation technique using fractional fourier
  transform for denoising medical images}.
\bjtitle{Optik}
\bvolume{124}(\bissue{3}),
\bfpage{265}--\blpage{271}
(\byear{2013})
\doiurl{10.1016/j.ijleo.2011.11.052}
\end{barticle}
\endbibitem

\bibitem[\protect\citeauthoryear{Zhang et~al.}{2021}]{Zhang2021}
\begin{bchapter}
\bauthor{\bsnm{Zhang}, \binits{X.}},
\bauthor{\bsnm{Liu}, \binits{Z.}},
\bauthor{\bsnm{Wang}, \binits{W.}},
\bauthor{\bsnm{Xu}, \binits{J.}}:
\bctitle{Automated detection of marine mammal species based on short-time
  fractional fourier transform}.
In: \bbtitle{{OCEANS} 2021: San Diego {\textendash} Porto}.
\bpublisher{{IEEE}}, \blocation{???}
(\byear{2021}).
\doiurl{10.23919/oceans44145.2021.9705945} .
\burl{https://doi.org/10.23919/oceans44145.2021.9705945}
\end{bchapter}
\endbibitem

\bibitem[\protect\citeauthoryear{Capus and Brown}{2003}]{Capus2003}
\begin{barticle}
\bauthor{\bsnm{Capus}, \binits{C.}},
\bauthor{\bsnm{Brown}, \binits{K.}}:
\batitle{Short-time fractional fourier methods for the time-frequency
  representation of chirp signals}.
\bjtitle{The Journal of the Acoustical Society of America}
\bvolume{113}(\bissue{6}),
\bfpage{3253}
(\byear{2003})
\doiurl{10.1121/1.1570434}
\end{barticle}
\endbibitem

\bibitem[\protect\citeauthoryear{Gorbunov and Dolovova}{2022}]{Gorbunov2022}
\begin{barticle}
\bauthor{\bsnm{Gorbunov}, \binits{M.}},
\bauthor{\bsnm{Dolovova}, \binits{O.}}:
\batitle{Fractional fourier transform and distributions in the ray space:
  Application for the analysis of radio occultation data}.
\bjtitle{Remote Sensing}
\bvolume{14}(\bissue{22}),
\bfpage{5802}
(\byear{2022})
\doiurl{10.3390/rs14225802}
\end{barticle}
\endbibitem

\bibitem[\protect\citeauthoryear{Huang and Shen}{2022}]{Huang2022}
\begin{barticle}
\bauthor{\bsnm{Huang}, \binits{L.}},
\bauthor{\bsnm{Shen}, \binits{X.}}:
\batitle{Research on speech emotion recognition based on the fractional fourier
  transform}.
\bjtitle{Electronics}
\bvolume{11}(\bissue{20}),
\bfpage{3393}
(\byear{2022})
\doiurl{10.3390/electronics11203393}
\end{barticle}
\endbibitem

\bibitem[\protect\citeauthoryear{Tian}{2021}]{Tian2021}
\begin{barticle}
\bauthor{\bsnm{Tian}, \binits{L.}}:
\batitle{Seismic spectral decomposition using short-time fractional fourier
  transform spectrograms}.
\bjtitle{Journal of Applied Geophysics}
\bvolume{192},
\bfpage{104400}
(\byear{2021})
\doiurl{10.1016/j.jappgeo.2021.104400}
\end{barticle}
\endbibitem

\bibitem[\protect\citeauthoryear{Cowell and
  Freear}{2010}]{cowell2010separation}
\begin{barticle}
\bauthor{\bsnm{Cowell}, \binits{D.M.}},
\bauthor{\bsnm{Freear}, \binits{S.}}:
\batitle{Separation of overlapping linear frequency modulated (lfm) signals
  using the fractional fourier transform}.
\bjtitle{IEEE transactions on ultrasonics, ferroelectrics, and frequency
  control}
\bvolume{57}(\bissue{10}),
\bfpage{2324}--\blpage{2333}
(\byear{2010})
\end{barticle}
\endbibitem

\bibitem[\protect\citeauthoryear{Tao et~al.}{2009}]{tao2009short}
\begin{barticle}
\bauthor{\bsnm{Tao}, \binits{R.}},
\bauthor{\bsnm{Li}, \binits{Y.-L.}},
\bauthor{\bsnm{Wang}, \binits{Y.}}:
\batitle{Short-time fractional fourier transform and its applications}.
\bjtitle{IEEE Transactions on Signal Processing}
\bvolume{58}(\bissue{5}),
\bfpage{2568}--\blpage{2580}
(\byear{2009})
\end{barticle}
\endbibitem

\bibitem[\protect\citeauthoryear{Zalevsky
  et~al.}{2007}]{zalevsky2007fractional}
\begin{barticle}
\bauthor{\bsnm{Zalevsky}, \binits{Z.}},
\bauthor{\bsnm{Ozaktas}, \binits{H.}},
\bauthor{\bsnm{Kutay}, \binits{A.}}:
\batitle{Fractional fourier transform-exceeding the classical concepts of
  signal’s manipulation}.
\bjtitle{Optics and Spectroscopy}
\bvolume{103},
\bfpage{868}--\blpage{876}
(\byear{2007})
\end{barticle}
\endbibitem

\bibitem[\protect\citeauthoryear{Boashash}{2015}]{boashash2015time}
\begin{bbook}
\bauthor{\bsnm{Boashash}, \binits{B.}}:
\bbtitle{Time-frequency Signal Analysis and Processing: a Comprehensive
  Reference}.
\bpublisher{Academic press}, \blocation{???}
(\byear{2015})
\end{bbook}
\endbibitem

\bibitem[\protect\citeauthoryear{Agrawal et~al.}{2023}]{Agrawal2023}
\begin{barticle}
\bauthor{\bsnm{Agrawal}, \binits{J.}},
\bauthor{\bsnm{Gupta}, \binits{M.}},
\bauthor{\bsnm{Garg}, \binits{H.}}:
\batitle{A review on speech separation in cocktail party environment:
  challenges and approaches}.
\bjtitle{Multimedia Tools and Applications}
(\byear{2023})
\doiurl{10.1007/s11042-023-14649-x}
\end{barticle}
\endbibitem

\bibitem[\protect\citeauthoryear{Novoselov et~al.}{2022}]{Novoselov2022}
\begin{botherref}
\oauthor{\bsnm{Novoselov}, \binits{A.}},
\oauthor{\bsnm{Balazs}, \binits{P.}},
\oauthor{\bsnm{Bokelmann}, \binits{G.}}:
{SEDENOSS}: {SEparating} and {DENOising} seismic signals with dual-path
  recurrent neural network architecture.
Journal of Geophysical Research: Solid Earth
\textbf{127}(3)
(2022)
\doiurl{10.1029/2021jb023183}
\end{botherref}
\endbibitem

\bibitem[\protect\citeauthoryear{Pfeifenberger and
  Pernkopf}{2022}]{Pfeifenberger2022}
\begin{barticle}
\bauthor{\bsnm{Pfeifenberger}, \binits{L.}},
\bauthor{\bsnm{Pernkopf}, \binits{F.}}:
\batitle{Blind speech separation and dereverberation using neural beamforming}.
\bjtitle{Speech Communication}
\bvolume{140},
\bfpage{29}--\blpage{41}
(\byear{2022})
\doiurl{10.1016/j.specom.2022.03.004}
\end{barticle}
\endbibitem

\bibitem[\protect\citeauthoryear{Gkalinikis
  et~al.}{2022}]{VirtsionisGkalinikis2022}
\begin{barticle}
\bauthor{\bsnm{Gkalinikis}, \binits{N.V.}},
\bauthor{\bsnm{Nalmpantis}, \binits{C.}},
\bauthor{\bsnm{Vrakas}, \binits{D.}}:
\batitle{Torch-{NILM}: An effective deep learning toolkit for non-intrusive
  load monitoring in pytorch}.
\bjtitle{Energies}
\bvolume{15}(\bissue{7}),
\bfpage{2647}
(\byear{2022})
\doiurl{10.3390/en15072647}
\end{barticle}
\endbibitem

\bibitem[\protect\citeauthoryear{Sun et~al.}{2020}]{Sun2020}
\begin{barticle}
\bauthor{\bsnm{Sun}, \binits{X.}},
\bauthor{\bsnm{Xu}, \binits{J.}},
\bauthor{\bsnm{Ma}, \binits{Y.}},
\bauthor{\bsnm{Zhao}, \binits{T.}},
\bauthor{\bsnm{Ou}, \binits{S.}},
\bauthor{\bsnm{Peng}, \binits{L.}}:
\batitle{Blind image separation based on attentional generative adversarial
  network}.
\bjtitle{Journal of Ambient Intelligence and Humanized Computing}
\bvolume{13}(\bissue{3}),
\bfpage{1397}--\blpage{1404}
(\byear{2020})
\doiurl{10.1007/s12652-020-02637-0}
\end{barticle}
\endbibitem

\bibitem[\protect\citeauthoryear{Wang et~al.}{2023}]{WANG2023104180}
\begin{barticle}
\bauthor{\bsnm{Wang}, \binits{W.}},
\bauthor{\bsnm{Wang}, \binits{S.}},
\bauthor{\bsnm{Qin}, \binits{D.}},
\bauthor{\bsnm{Fang}, \binits{Y.}},
\bauthor{\bsnm{Zheng}, \binits{Y.}}:
\batitle{Heart-lung sound separation by nonnegative matrix factorization and
  deep learning}.
\bjtitle{Biomedical Signal Processing and Control}
\bvolume{79},
\bfpage{104180}
(\byear{2023})
\doiurl{10.1016/j.bspc.2022.104180}
\end{barticle}
\endbibitem

\bibitem[\protect\citeauthoryear{Karhunen
  et~al.}{1997}]{Karhunen_1997_NN_for_ICA}
\begin{barticle}
\bauthor{\bsnm{Karhunen}, \binits{J.}},
\bauthor{\bsnm{Oja}, \binits{E.}},
\bauthor{\bsnm{Wang}, \binits{L.}},
\bauthor{\bsnm{Vigario}, \binits{R.}},
\bauthor{\bsnm{Joutsensalo}, \binits{J.}}:
\batitle{A class of neural networks for independent component analysis}.
\bjtitle{IEEE Transactions on Neural Networks}
\bvolume{8}(\bissue{3}),
\bfpage{486}--\blpage{504}
(\byear{1997})
\doiurl{10.1109/72.572090}
\end{barticle}
\endbibitem

\bibitem[\protect\citeauthoryear{Bermant}{2021}]{Bermant2021}
\begin{botherref}
\oauthor{\bsnm{Bermant}, \binits{P.C.}}:
{BioCPPNet}: automatic bioacoustic source separation with deep neural networks.
Scientific Reports
\textbf{11}(1)
(2021)
\doiurl{10.1038/s41598-021-02790-2}
\end{botherref}
\endbibitem

\bibitem[\protect\citeauthoryear{Chen et~al.}{2022}]{chen2022survey}
\begin{barticle}
\bauthor{\bsnm{Chen}, \binits{W.-S.}},
\bauthor{\bsnm{Zeng}, \binits{Q.}},
\bauthor{\bsnm{Pan}, \binits{B.}}:
\batitle{A survey of deep nonnegative matrix factorization}.
\bjtitle{Neurocomputing}
\bvolume{491},
\bfpage{305}--\blpage{320}
(\byear{2022})
\end{barticle}
\endbibitem

\bibitem[\protect\citeauthoryear{Nakamura et~al.}{2021}]{Nakamura_2021}
\begin{barticle}
\bauthor{\bsnm{Nakamura}, \binits{T.}},
\bauthor{\bsnm{Kozuka}, \binits{S.}},
\bauthor{\bsnm{Saruwatari}, \binits{H.}}:
\batitle{Time-domain audio source separation with neural networks based on
  multiresolution analysis}.
\bjtitle{IEEE/ACM Transactions on Audio, Speech, and Language Processing}
\bvolume{29},
\bfpage{1687}--\blpage{1701}
(\byear{2021})
\doiurl{10.1109/TASLP.2021.3072496}
\end{barticle}
\endbibitem

\bibitem[\protect\citeauthoryear{Dai et~al.}{2021}]{Dai2021}
\begin{barticle}
\bauthor{\bsnm{Dai}, \binits{Y.}},
\bauthor{\bsnm{Yang}, \binits{J.}},
\bauthor{\bsnm{Dong}, \binits{Y.}},
\bauthor{\bsnm{Zou}, \binits{H.}},
\bauthor{\bsnm{Hu}, \binits{M.}},
\bauthor{\bsnm{Wang}, \binits{B.}}:
\batitle{Blind source separation-based {IVA}-xception model for bird sound
  recognition in complex acoustic environments}.
\bjtitle{Electronics Letters}
\bvolume{57}(\bissue{11}),
\bfpage{454}--\blpage{456}
(\byear{2021})
\doiurl{10.1049/ell2.12160}
\end{barticle}
\endbibitem

\bibitem[\protect\citeauthoryear{Chatterji et~al.}{2004}]{Q_transform}
\begin{barticle}
\bauthor{\bsnm{Chatterji}, \binits{S.}},
\bauthor{\bsnm{Blackburn}, \binits{L.}},
\bauthor{\bsnm{Martin}, \binits{G.}},
\bauthor{\bsnm{Katsavounidis}, \binits{E.}}:
\batitle{Multiresolution techniques for the detection of gravitational-wave
  bursts}.
\bjtitle{Classical and Quantum Gravity}
\bvolume{21}(\bissue{20}),
\bfpage{1809}--\blpage{1818}
(\byear{2004})
\doiurl{10.1088/0264-9381/21/20/024}
\end{barticle}
\endbibitem

\bibitem[\protect\citeauthoryear{Nitz et~al.}{2023}]{pyCBC}
\begin{botherref}
\oauthor{\bsnm{Nitz}, \binits{A.}}, et al.:
pyCBC
(2023).
\doiurl{10.5281/zenodo.7885796}
\end{botherref}
\endbibitem

\bibitem[\protect\citeauthoryear{Chassande-Mottin and
  Flandrin}{1999}]{CHASSANDEMOTTIN1999252}
\begin{barticle}
\bauthor{\bsnm{Chassande-Mottin}, \binits{E.}},
\bauthor{\bsnm{Flandrin}, \binits{P.}}:
\batitle{On the time–frequency detection of chirps}.
\bjtitle{Applied and Computational Harmonic Analysis}
\bvolume{6}(\bissue{2}),
\bfpage{252}--\blpage{281}
(\byear{1999})
\doiurl{10.1006/acha.1998.0254}
\end{barticle}
\endbibitem

\bibitem[\protect\citeauthoryear{Moore et~al.}{2014}]{Moore2014}
\begin{barticle}
\bauthor{\bsnm{Moore}, \binits{C.J.}},
\bauthor{\bsnm{Cole}, \binits{R.H.}},
\bauthor{\bsnm{Berry}, \binits{C.P.L.}}:
\batitle{Gravitational-wave sensitivity curves}.
\bjtitle{Classical and Quantum Gravity}
\bvolume{32}(\bissue{1}),
\bfpage{015014}
(\byear{2014})
\doiurl{10.1088/0264-9381/32/1/015014}
\end{barticle}
\endbibitem

\bibitem[\protect\citeauthoryear{Davis et~al.}{2022}]{Davis2022}
\begin{barticle}
\bauthor{\bsnm{Davis}, \binits{D.}},
\bauthor{\bsnm{Littenberg}, \binits{T.B.}},
\bauthor{\bsnm{Romero-Shaw}, \binits{I.M.}},
\bauthor{\bsnm{Millhouse}, \binits{M.}},
\bauthor{\bsnm{McIver}, \binits{J.}},
\bauthor{\bsnm{Renzo}, \binits{F.D.}},
\bauthor{\bsnm{Ashton}, \binits{G.}}:
\batitle{Subtracting glitches from gravitational-wave detector data during the
  third {LIGO}-virgo observing run}.
\bjtitle{Classical and Quantum Gravity}
\bvolume{39}(\bissue{24}),
\bfpage{245013}
(\byear{2022})
\doiurl{10.1088/1361-6382/aca238}
\end{barticle}
\endbibitem

\bibitem[\protect\citeauthoryear{Maggiore et~al.}{2020}]{Maggiore2020}
\begin{barticle}
\bauthor{\bsnm{Maggiore}, \binits{M.}},
\bauthor{\bsnm{Broeck}, \binits{C.V.D.}},
\bauthor{\bsnm{Bartolo}, \binits{N.}},
\bauthor{\bsnm{Belgacem}, \binits{E.}},
\bauthor{\bsnm{Bertacca}, \binits{D.}},
\bauthor{\bsnm{Bizouard}, \binits{M.A.}},
\bauthor{\bsnm{Branchesi}, \binits{M.}},
\bauthor{\bsnm{Clesse}, \binits{S.}},
\bauthor{\bsnm{Foffa}, \binits{S.}},
\bauthor{\bsnm{Garc{\'{\i}}a-Bellido}, \binits{J.}},
\bauthor{\bsnm{Grimm}, \binits{S.}},
\bauthor{\bsnm{Harms}, \binits{J.}},
\bauthor{\bsnm{Hinderer}, \binits{T.}},
\bauthor{\bsnm{Matarrese}, \binits{S.}},
\bauthor{\bsnm{Palomba}, \binits{C.}},
\bauthor{\bsnm{Peloso}, \binits{M.}},
\bauthor{\bsnm{Ricciardone}, \binits{A.}},
\bauthor{\bsnm{Sakellariadou}, \binits{M.}}:
\batitle{Science case for the einstein telescope}.
\bjtitle{Journal of Cosmology and Astroparticle Physics}
\bvolume{2020}(\bissue{03}),
\bfpage{050}--\blpage{050}
(\byear{2020})
\doiurl{10.1088/1475-7516/2020/03/050}
\end{barticle}
\endbibitem

\bibitem[\protect\citeauthoryear{Crowder and Cornish}{2004}]{Crowder2004}
\begin{botherref}
\oauthor{\bsnm{Crowder}, \binits{J.}},
\oauthor{\bsnm{Cornish}, \binits{N.J.}}:
{LISA} source confusion.
Physical Review D
\textbf{70}(8)
(2004)
\doiurl{10.1103/physrevd.70.082004}
\end{botherref}
\endbibitem

\bibitem[\protect\citeauthoryear{Littenberg et~al.}{2020}]{Littenberg2020}
\begin{botherref}
\oauthor{\bsnm{Littenberg}, \binits{T.B.}},
\oauthor{\bsnm{Cornish}, \binits{N.J.}},
\oauthor{\bsnm{Lackeos}, \binits{K.}},
\oauthor{\bsnm{Robson}, \binits{T.}}:
Global analysis of the gravitational wave signal from galactic binaries.
Physical Review D
\textbf{101}(12)
(2020)
\doiurl{10.1103/physrevd.101.123021}
\end{botherref}
\endbibitem

\bibitem[\protect\citeauthoryear{Messick et~al.}{2017}]{GstLAL}
\begin{botherref}
\oauthor{\bsnm{Messick}, \binits{C.}},
\oauthor{\bsnm{Blackburn}, \binits{K.}},
\oauthor{\bsnm{Brady}, \binits{P.}},
\oauthor{\bsnm{Brockill}, \binits{P.}},
\oauthor{\bsnm{Cannon}, \binits{K.}},
\oauthor{\bsnm{Cariou}, \binits{R.}},
\oauthor{\bsnm{Caudill}, \binits{S.}},
\oauthor{\bsnm{Chamberlin}, \binits{S.J.}},
\oauthor{\bsnm{Creighton}, \binits{J.D.E.}},
\oauthor{\bsnm{Everett}, \binits{R.}},
\oauthor{\bsnm{Hanna}, \binits{C.}},
\oauthor{\bsnm{Keppel}, \binits{D.}},
\oauthor{\bsnm{Lang}, \binits{R.N.}},
\oauthor{\bsnm{Li}, \binits{T.G.F.}},
\oauthor{\bsnm{Meacher}, \binits{D.}},
\oauthor{\bsnm{Nielsen}, \binits{A.}},
\oauthor{\bsnm{Pankow}, \binits{C.}},
\oauthor{\bsnm{Privitera}, \binits{S.}},
\oauthor{\bsnm{Qi}, \binits{H.}},
\oauthor{\bsnm{Sachdev}, \binits{S.}},
\oauthor{\bsnm{Sadeghian}, \binits{L.}},
\oauthor{\bsnm{Singer}, \binits{L.}},
\oauthor{\bsnm{Thomas}, \binits{E.G.}},
\oauthor{\bsnm{Wade}, \binits{L.}},
\oauthor{\bsnm{Wade}, \binits{M.}},
\oauthor{\bsnm{Weinstein}, \binits{A.}},
\oauthor{\bsnm{Wiesner}, \binits{K.}}:
Analysis framework for the prompt discovery of compact binary mergers in
  gravitational-wave data.
Physical Review D
\textbf{95}(4)
(2017)
\doiurl{10.1103/physrevd.95.042001}
\end{botherref}
\endbibitem

\bibitem[\protect\citeauthoryear{Aubin et~al.}{2021}]{MBTA}
\begin{barticle}
\bauthor{\bsnm{Aubin}, \binits{F.}},
\bauthor{\bsnm{Brighenti}, \binits{F.}},
\bauthor{\bsnm{Chierici}, \binits{R.}},
\bauthor{\bsnm{Estevez}, \binits{D.}},
\bauthor{\bsnm{Greco}, \binits{G.}},
\bauthor{\bsnm{Guidi}, \binits{G.M.}},
\bauthor{\bsnm{Juste}, \binits{V.}},
\bauthor{\bsnm{Marion}, \binits{F.}},
\bauthor{\bsnm{Mours}, \binits{B.}},
\bauthor{\bsnm{Nitoglia}, \binits{E.}},
\bauthor{\bsnm{Sauter}, \binits{O.}},
\bauthor{\bsnm{Sordini}, \binits{V.}}:
\batitle{The {MBTA} pipeline for detecting compact binary coalescences in the
  third {LIGO}{\textendash}virgo observing run}.
\bjtitle{Classical and Quantum Gravity}
\bvolume{38}(\bissue{9}),
\bfpage{095004}
(\byear{2021})
\doiurl{10.1088/1361-6382/abe913}
\end{barticle}
\endbibitem

\bibitem[\protect\citeauthoryear{Nitz et~al.}{2018}]{PyCBC_LIVE}
\begin{botherref}
\oauthor{\bsnm{Nitz}, \binits{A.H.}},
\oauthor{\bsnm{Canton}, \binits{T.D.}},
\oauthor{\bsnm{Davis}, \binits{D.}},
\oauthor{\bsnm{Reyes}, \binits{S.}}:
Rapid detection of gravitational waves from compact binary mergers with {PyCBC}
  live.
Physical Review D
\textbf{98}(2)
(2018)
\doiurl{10.1103/physrevd.98.024050}
\end{botherref}
\endbibitem

\bibitem[\protect\citeauthoryear{Hooper et~al.}{2012}]{SPIIR}
\begin{botherref}
\oauthor{\bsnm{Hooper}, \binits{S.}},
\oauthor{\bsnm{Chung}, \binits{S.K.}},
\oauthor{\bsnm{Luan}, \binits{J.}},
\oauthor{\bsnm{Blair}, \binits{D.}},
\oauthor{\bsnm{Chen}, \binits{Y.}},
\oauthor{\bsnm{Wen}, \binits{L.}}:
Summed parallel infinite impulse response filters for low-latency detection of
  chirping gravitational waves.
Physical Review D
\textbf{86}(2)
(2012)
\doiurl{10.1103/physrevd.86.024012}
\end{botherref}
\endbibitem

\bibitem[\protect\citeauthoryear{Klimenko et~al.}{2016}]{cWB}
\begin{botherref}
\oauthor{\bsnm{Klimenko}, \binits{S.}},
\oauthor{\bsnm{Vedovato}, \binits{G.}},
\oauthor{\bsnm{Drago}, \binits{M.}},
\oauthor{\bsnm{Salemi}, \binits{F.}},
\oauthor{\bsnm{Tiwari}, \binits{V.}},
\oauthor{\bsnm{Prodi}, \binits{G.A.}},
\oauthor{\bsnm{Lazzaro}, \binits{C.}},
\oauthor{\bsnm{Ackley}, \binits{K.}},
\oauthor{\bsnm{Tiwari}, \binits{S.}},
\oauthor{\bsnm{Silva}, \binits{C.F.D.}},
\oauthor{\bsnm{Mitselmakher}, \binits{G.}}:
Method for detection and reconstruction of gravitational wave transients with
  networks of advanced detectors.
Physical Review D
\textbf{93}(4)
(2016)
\doiurl{10.1103/physrevd.93.042004}
\end{botherref}
\endbibitem

\bibitem[\protect\citeauthoryear{Lynch et~al.}{2017}]{oLIB}
\begin{botherref}
\oauthor{\bsnm{Lynch}, \binits{R.}},
\oauthor{\bsnm{Vitale}, \binits{S.}},
\oauthor{\bsnm{Essick}, \binits{R.}},
\oauthor{\bsnm{Katsavounidis}, \binits{E.}},
\oauthor{\bsnm{Robinet}, \binits{F.}}:
Information-theoretic approach to the gravitational-wave burst detection
  problem.
Physical Review D
\textbf{95}(10)
(2017)
\doiurl{10.1103/physrevd.95.104046}
\end{botherref}
\endbibitem

\bibitem[\protect\citeauthoryear{Cornish and Littenberg}{2015}]{BayesWave}
\begin{barticle}
\bauthor{\bsnm{Cornish}, \binits{N.J.}},
\bauthor{\bsnm{Littenberg}, \binits{T.B.}}:
\batitle{Bayeswave: Bayesian inference for gravitational wave bursts and
  instrument glitches}.
\bjtitle{Classical and Quantum Gravity}
\bvolume{32}(\bissue{13}),
\bfpage{135012}
(\byear{2015})
\doiurl{10.1088/0264-9381/32/13/135012}
\end{barticle}
\endbibitem

\bibitem[\protect\citeauthoryear{Forte et~al.}{2011}]{Forte2011WignerVille}
\begin{botherref}
\oauthor{\bsnm{Forte}, \binits{L.A.}},
\oauthor{\bsnm{Garufi}, \binits{F.}},
\oauthor{\bsnm{Milano}, \binits{L.}},
\oauthor{\bsnm{Croce}, \binits{R.P.}},
\oauthor{\bsnm{Pierro}, \binits{V.}},
\oauthor{\bsnm{Pinto}, \binits{I.}}:
Blind source separation and wigner-ville transform as tools for the extraction
  of the gravitational wave signal.
Physical Review D
\textbf{83}(12)
(2011)
\doiurl{10.1103/physrevd.83.122006}
\end{botherref}
\endbibitem

\bibitem[\protect\citeauthoryear{Huang}{2005}]{Huang2005}
\begin{bchapter}
\bauthor{\bsnm{Huang}, \binits{N.E.}}:
\bctitle{{INTRODUCTION} {TO} {THE} {HILBERT}{\textendash}{HUANG} {TRANSFORM}
  {AND} {ITS} {RELATED} {MATHEMATICAL} {PROBLEMS}}.
In: \bbtitle{Hilbert-Huang Transform and Its Applications},
pp. \bfpage{1}--\blpage{26}.
\bpublisher{{WORLD} {SCIENTIFIC}}, \blocation{???}
(\byear{2005}).
\doiurl{10.1142/9789812703347_0001} .
\burl{https://doi.org/10.1142/9789812703347_0001}
\end{bchapter}
\endbibitem

\bibitem[\protect\citeauthoryear{Hu et~al.}{2022}]{Hu2022}
\begin{barticle}
\bauthor{\bsnm{Hu}, \binits{C.-P.}},
\bauthor{\bsnm{Lin}, \binits{L.C.-C.}},
\bauthor{\bsnm{Pan}, \binits{K.-C.}},
\bauthor{\bsnm{Li}, \binits{K.-L.}},
\bauthor{\bsnm{Yen}, \binits{C.-C.}},
\bauthor{\bsnm{Kong}, \binits{A.K.H.}},
\bauthor{\bsnm{Hui}, \binits{C.Y.}}:
\batitle{A comprehensive analysis of the gravitational wave events with the
  stacked hilbert{\textendash}huang transform: From compact binary coalescence
  to supernova}.
\bjtitle{The Astrophysical Journal}
\bvolume{935}(\bissue{2}),
\bfpage{127}
(\byear{2022})
\doiurl{10.3847/1538-4357/ac8165}
\end{barticle}
\endbibitem

\bibitem[\protect\citeauthoryear{Debnath}{2002}]{Debnath2002}
\begin{bbook}
\bauthor{\bsnm{Debnath}, \binits{L.}}:
\bbtitle{The Wigner-Ville Distribution and Time-Frequency Signal Analysis},
pp. \bfpage{307}--\blpage{360}.
\bpublisher{Birkh{\"a}user Boston},
\blocation{Boston, MA}
(\byear{2002}).
\doiurl{10.1007/978-1-4612-0097-0_5} .
\burl{https://doi.org/10.1007/978-1-4612-0097-0_5}
\end{bbook}
\endbibitem

\bibitem[\protect\citeauthoryear{Abbott et~al.}{2017}]{Abbott2017}
\begin{botherref}
\oauthor{\bsnm{Abbott}, \binits{B.P.}}, et al.:
{GW}170817: Observation of gravitational waves from a binary neutron star
  inspiral.
Physical Review Letters
\textbf{119}(16)
(2017)
\doiurl{10.1103/physrevlett.119.161101}
\end{botherref}
\endbibitem

\bibitem[\protect\citeauthoryear{Harms et~al.}{2021}]{Harms2021}
\begin{barticle}
\bauthor{\bsnm{Harms}, \binits{J.}}, \betal:
\batitle{Lunar gravitational-wave antenna}.
\bjtitle{The Astrophysical Journal}
\bvolume{910}(\bissue{1}),
\bfpage{1}
(\byear{2021})
\doiurl{10.3847/1538-4357/abe5a7}
\end{barticle}
\endbibitem

\end{thebibliography}

\end{document}